\documentclass[apj]{emulateapj}

\usepackage{graphicx}
\usepackage{amssymb}
\usepackage{amsmath}    
\usepackage{txfonts}
\usepackage{xspace}
\usepackage{natbib}
\bibpunct{(}{)}{;}{a}{}{,}

\newcommand{\Chandra}{\textit{Chandra}\xspace}
\newcommand{\Fermi}{\textit{Fermi}\xspace}
\newcommand{\FermiLAT}{\textit{Fermi}-LAT\xspace}
\newcommand{\HST}{\textit{HST}\xspace}
\newcommand{\He}{H.E.S.S.\xspace}
\newcommand{\Ve}{VERITAS\xspace}

\newcommand{\Ma}{MAGIC\xspace}

\newcommand{\M}{M\,87\xspace}

\newcommand{\degree}{\hbox{$^\circ$}\xspace}

\shorttitle{2010 VHE flare \& 10 years of MWL observations of \M}
\shortauthors{The \He, \Ma, \& \Ve Collaborations and the \M MWL Monitoring Team}

\begin{document}


\title{The 2010 very high energy $\gamma$-ray  flare \& 10 years of multi-wavelength observations of \M}


\author{
The H.E.S.S. Collaboration:
A.~Abramowski\altaffilmark{1},
F.~Acero\altaffilmark{2},
F.~Aharonian\altaffilmark{3,4,5},
A.G.~Akhperjanian\altaffilmark{6,5},
G.~Anton\altaffilmark{7},
A.~Balzer\altaffilmark{7},
A.~Barnacka\altaffilmark{8,9},
U.~Barres~de~Almeida\altaffilmark{10},
Y.~Becherini\altaffilmark{11,12}
% ...
}

% ...
\altaffiltext{1}{Universit\"at Hamburg, Institut f\"ur Experimentalphysik, Luruper Chaussee 149, D 22761 Hamburg, Germany}
\altaffiltext{2}{Laboratoire Univers et Particules de Montpellier, Universit\'e Montpellier 2, CNRS/IN2P3,  CC 72, Place Eug\`ene Bataillon, F-34095 Montpellier Cedex 5, France}
\altaffiltext{3}{Max-Planck-Institut f\"ur Kernphysik, P.O. Box 103980, D 69029 Heidelberg, Germany}
\altaffiltext{4}{Dublin Institute for Advanced Studies, 31 Fitzwilliam Place, Dublin 2, Ireland}
\altaffiltext{5}{National Academy of Sciences of the Republic of Armenia, Yerevan}
\altaffiltext{6}{Yerevan Physics Institute, 2 Alikhanian Brothers St., 375036 Yerevan, Armenia}
\altaffiltext{7}{Universit\"at Erlangen-N\"urnberg, Physikalisches Institut, Erwin-Rommel-Str. 1, D 91058 Erlangen, Germany}
\altaffiltext{8}{Nicolaus Copernicus Astronomical Center, ul. Bartycka 18, 00-716 Warsaw, Poland}
\altaffiltext{9}{CEA Saclay, DSM/IRFU, F-91191 Gif-Sur-Yvette Cedex, France}
\altaffiltext{10}{University of Durham, Department of Physics, South Road, Durham DH1 3LE, U.K.}
\altaffiltext{11}{Astroparticule et Cosmologie (APC), CNRS, Universit\'{e} Paris 7 Denis Diderot, 10, rue Alice Domon et L\'{e}onie Duquet, F-75205 Paris Cedex 13, France \thanks{(UMR 7164: CNRS, Universit\'e Paris VII, CEA, Observatoire de Paris)}}
\altaffiltext{12}{Laboratoire Leprince-Ringuet, Ecole Polytechnique, CNRS/IN2P3, F-91128 Palaiseau, France}
% ...




\begin{abstract}
The giant radio galaxy \M with its proximity (16\,Mpc), famous jet, and very massive black hole ($(3 - 6) \times10^9$M$_\odot$) provides a unique opportunity to investigate the origin of very high energy (VHE; E$>$100 GeV) $\gamma$-ray emission  generated in relativistic outflows and the surroundings of super-massive black holes.
 \M has been established as a VHE $\gamma$-ray emitter since 2006. The VHE $\gamma$-ray emission displays strong variability on timescales as short as a day.
In this paper, results from a joint VHE monitoring campaign on \M by the \Ma and \Ve instruments in 2010 are reported. During the campaign, a flare at VHE was detected triggering further observations at VHE (\He), X-rays (\Chandra), and radio (43\,GHz VLBA). The excellent sampling of the VHE $\gamma$-ray light curve enables one to derive a precise temporal characterization of the flare: the single, isolated flare is well described by a two-sided exponential function with significantly different flux rise and decay times of $\tau_\text{d}^{\text{rise}} = (1.69 \pm 0.30)$\,days and $\tau_\text{d}^{\text{decay}} = (0.611 \pm 0.080)$\,days, respectively.
While the overall variability pattern of the 2010 flare appears somewhat different from that of previous VHE flares in 2005 and 2008, they share very similar timescales ($\sim$day), peak fluxes ($\Phi_{>0.35\,{\rm TeV}} \simeq (1-3) \times 10^{-11}$\,ph\,cm$^{-2}$\,s$^{-1}$), and VHE spectra.
43\,GHz VLBA radio observations of the inner jet regions indicate no enhanced flux in 2010 in contrast to observations in 2008, where an increase of the radio flux of the innermost core regions coincided with a VHE flare. On the other hand, \Chandra X-ray observations taken $\sim3$\,days after the peak of the VHE $\gamma$-ray emission reveal an enhanced flux from the core (flux increased by factor $\sim2$; variability timescale $<2$ days).
The long-term (2001-2010) multi-wavelength (MWL) light curve of \M, spanning from radio to VHE and including data from \HST, LT, VLA and EVN, is used to further investigate the origin of the VHE $\gamma$-ray emission.
No unique, common MWL signature of the three VHE flares has been identified.
In the outer kpc jet region, in particular in HST-1, no enhanced MWL activity was detected in 2008 and 2010, disfavoring it as the origin of the VHE flares during these years.
Shortly after two of the three flares (2008 and 2010), the X-ray core was observed to be at a higher flux level than its characteristic range (determined from more than 60 monitoring observations: 2002-2009). In 2005, the strong flux dominance of HST-1 could have suppressed the detection of such a feature.
Published models for VHE $\gamma$-ray emission from \M are reviewed in the light of the new data.
 \end{abstract}



\section{Introduction}

The giant radio galaxy \M provides a unique environment to study relativistic plasma outflows and the surroundings of super-massive black holes (SMBH). Its prominent jet \citep{curtis:1918a:m87jet} is resolved from radio to X-rays displaying complex structures (knots, diffuse emission; \citealt{perlman:1999a,perlman:2001a}), strong variability \citep{harris:2003a,harris:2006a}, and apparent super-luminal motion \citep{biretta:1999a,cheung:2007a}. With its proximity ($16.7 \pm 0.2$\,Mpc; \citealt{mei:2007a}) and its very massive black hole of $M_{\rm BH} \simeq (3 - 6) \times 10^{9}$\,M$_\odot$\footnote{In the following $M_{\rm BH} \simeq 3 \times 10^{9}$\,M$_\odot$ is adopted.} \citep{macchetto:1997a,gebhardt:2009a}
high resolution very long baseline interferometry (VLBI) at radio wavelengths enables one to directly probe structures with sizes down to $<200$ Schwarzschild radii. From the detection of super-luminal features in the jet in the optical and radio the jet orientation angle towards the line of sight at the sub-kpc scale is limited to $\theta \lesssim 20$\degree \citep{biretta:1999a,cheung:2007a}.

Evidence for very high energy (VHE; $E>100$\,GeV) $\gamma$-ray emission from \M was reported by the HEGRA collaboration in 2003 \citep{aharonian:2003b}\footnote{See also \citet{lebohec:2004:whipple:m87}.} and was later confirmed by H.E.S.S., VERITAS, and MAGIC \citep{aharonian:2006:hess:m87:science,acciari:2008a:veritas:m87,albert:2008:magic:m87}. While the large majority of active galactic nuclei (AGN) with detected VHE $\gamma$-ray emission are strongly beamed sources, \M is one of only four known VHE AGN with weak or, at most, moderate beaming; the other three being the radio galaxies Centaurus\,A \citep{aharonian:2009:hess:cena}, IC\,310 \citep{aleksic:2011:magic:ic310}, and NGC\,1275 \citep{mariotti:2010:magic:ngc1275:atel}. Interestingly, for such a weakly beamed source, \M shows strong variability at VHE with timescales of the order of days \citep{aharonian:2006:hess:m87:science,albert:2008:magic:m87,acciari:2009b:m87joinedcampaign:science}. This points, through the causality argument, towards a compact emission region $< 5 \times 10^{15} \delta$\,cm ($\delta$ being the Doppler factor of the emitting plasma) corresponding to only a few Schwarzschild radii
$R_{\rm S} = 2 G M_{\rm BH} / c^2 \simeq 10^{15}$\,cm.
At GeV energies \M has recently been detected as a weak source by \FermiLAT \citep{abdo:2009:fermi:lat:m87}.

The exact location of the VHE $\gamma$-ray emitting region in \M remains elusive. The angular resolution of ground-based VHE instruments is of the order of 0.1\degree (corresponding to $\sim30$\,kpc projected size) and, therefore, does not allow for a direct precise determination of the VHE $\gamma$-ray emission site in the inner kpc-scale  structures, although the outer radio lobes can be excluded as the origin \citep{aharonian:2006:hess:m87:science}. To further investigate the location of the VHE $\gamma$-ray emission site and the associated production mechanisms, variability studies and the search for correlations with other wavelengths have successfully been utilized \citep[e.g.][]{acciari:2009b:m87joinedcampaign:science}. Of particular interest are radio observations, since they allow for the highest angular resolution, and X-ray observations, due to their potential connection with the VHE $\gamma$-ray emission in e.g. synchrotron self-Compton (SSC) models.

Up to now, three episodes of enhanced VHE $\gamma$-ray emission have been detected from \M, with details on the latest one, observed in 2010, being reported in this paper. The first one, detected in 2005 \citep{aharonian:2006:hess:m87:science}, coincided with an extreme multi-frequency outburst of the jet feature HST-1 \citep{harris:2003a,harris:2006a}.\footnote{The outburst was also followed by the ejection of apparent superluminal radio components from HST-1 \citep{cheung:2007a}.}, which has also been discussed as a possible VHE $\gamma$-ray emission site \citep[e.g.][]{stawarz:2006a,cheung:2007a,harris:2009a}. During the second flaring episode, detected in 2008, HST-1 was in a low flux state, but radio measurements at 43\,GHz with the Very Long Baseline Array (VLBA) showed a flux increase in the core region within a few hundred Schwarzschild radii of the SMBH, suggesting the direct vicinity of the SMBH as the origin of the VHE $\gamma$-ray emission \citep{acciari:2009b:m87joinedcampaign:science}. This conclusion was further supported by the detection of an enhanced X-ray flux from the core region by \Chandra.

In this paper results from a joint observation campaign of \M in 2010 are presented and discussed in the broader context of the multi-wavelength (MWL) behavior of \M over the past ten years. During the campaign, a high flux state at VHE was detected \citep{mariotti:2010:magic:m87:atel,ong:2010a:m87:veritas:magic:flare:atel}. Characteristics of the VHE flare are investigated and possible correlations with other wavelengths are discussed.
New observational results from \He, \Ma, \Ve, \FermiLAT, \Chandra, \HST, LT, VLBA, MOJAVE, VLA, and the EVN are presented.%
\footnote{Full names of the instruments can be found in Sec.~\ref{Sec:Data}.}

In Sec.~\ref{Sec:Data} the instruments and the data are introduced. In Sec.~\ref{Sec:2010VHECampaign} the characteristics of the VHE high state are investigated and compared to previous flares. New results from optical polarimetry observations with the Hubble Space Telescope and the Liverpool Telescope are presented and discussed in Sec.~\ref{Sec:OpticalPolarimetry}. In Sec.~\ref{Sec:VHEFlaresVsMWL} the VHE results are confronted with the broader MWL picture and the theoretical implications concerning the VHE $\gamma$-ray emission site are presented in Sec.~\ref{Sec:Discussion}. The paper concludes with Sec.~\ref{Sec:SummaryConclusions}.


\section{Data}\label{Sec:Data}

\begin{figure*}
\centering
\includegraphics[width=0.7\textwidth]{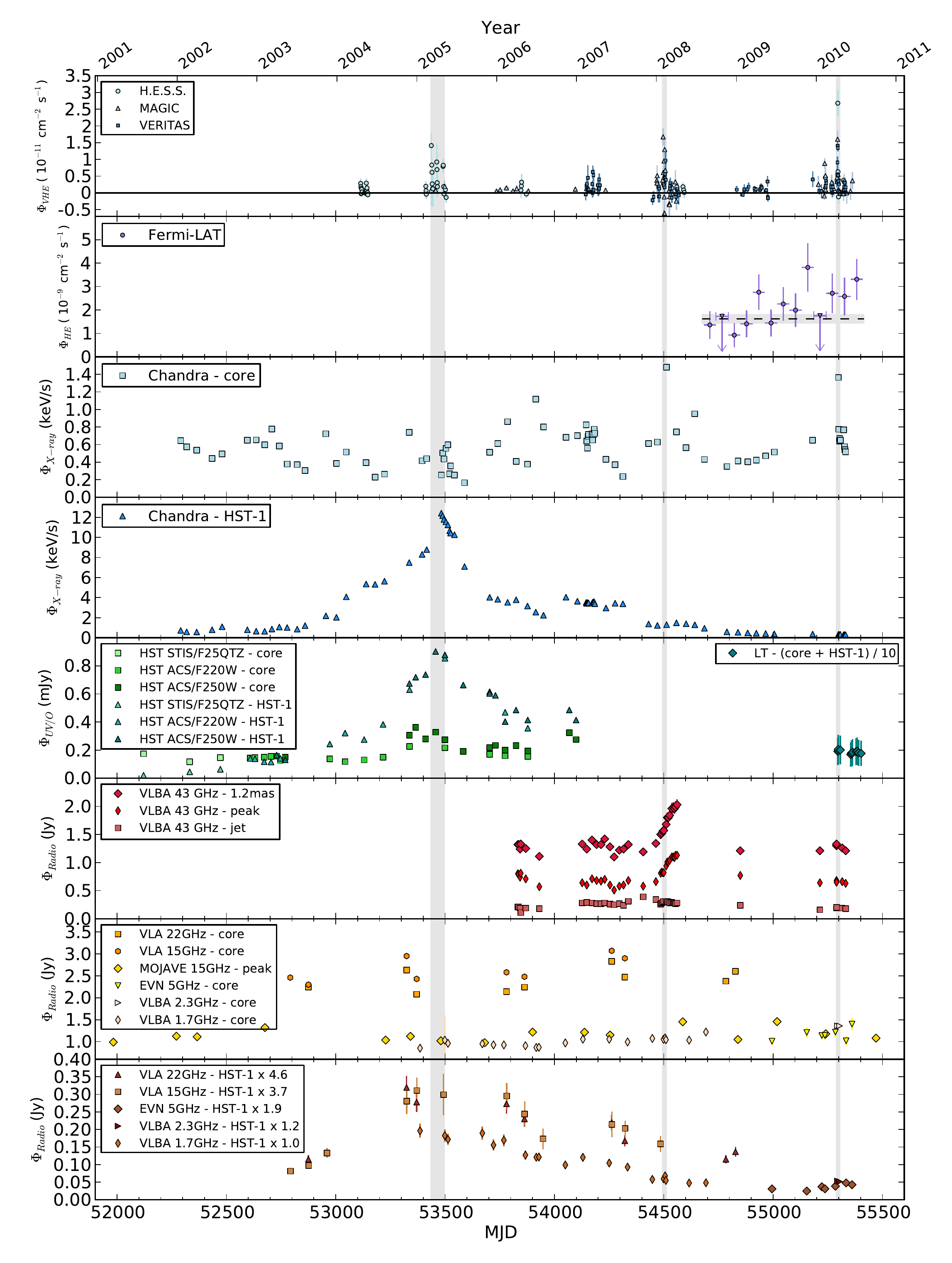}
\caption{Multi-wavelength light curve of \M from 2001 to 2011. The VHE $\gamma$-ray flux (top panel) is calculated above an energy threshold of 350\,GeV (see text). Separate fluxes for the core and HST-1 are shown in cases where the instrument resolution is sufficient to separate the two components. Gray vertical bands mark the times of increased VHE activity in 2005, 2008, and 2010 (see Fig.~\ref{Fig:VHELightcurves2005-10}). The dashed line and the gray horizontal band in the 2nd panel marks the average flux with 1\,s.d.\ error measured by \FermiLAT. The radio flux of HST-1 at different frequencies has been normalized to the 5\,GHz flux assuming a spectrum $S_\nu \sim \nu^{-\alpha}$
with $\alpha = 0.6$.
All flux errors shown are the 1\,s.d.\ statistical errors except for the LT data where the uncertainty on the contribution from the galaxy is included in the error bars. For details on the data, the data analysis, and references see text.}
\label{Fig:MWL-longterm-lightcurve}
\end{figure*}
 
 \subsection{Very high energy (VHE)}

\M has been the target of several coordinated VHE monitoring campaigns since 2008. In 2010 the source was jointly monitored by \Ma and \Ve. During the campaign \citet{mariotti:2010:magic:m87:atel} reported an increased flux from the source in February 2010, but follow-up observations did not reveal further activity at VHE. A second increase in flux, marking the onset of a strong VHE flare, was detected in April 2010 by \Ma and \Ve \citep{ong:2010a:m87:veritas:magic:flare:atel} triggering further ToO (Target of Opportunity) observations by \He, \Chandra, the VLBA, and the EVN.

For the 2010 campaign, \M has been observed for a total of $\sim$80\,h from December 2009 to June 2010. The data from different VHE instruments have been combined after separate analysis within the individual collaborations. Integral fluxes for the VHE band $\Phi_\text{VHE}$ are calculated above an energy of 350\,GeV. Observations taken or published with a different energy threshold are extrapolated using the average measured VHE spectrum, which is well described by a power law spectrum $\text{d}N/\text{d}E \propto E^{-\Gamma}$ with photon index $\Gamma = 2.3 \pm 0.11$ \citep{albert:2008:magic:m87}.

Given the indications for spectral variability of \M at VHE \citep{aharonian:2006:hess:m87:science,albert:2008:magic:m87,aliu:2011a:veritas:m87:2010} using a single photon index for the flux extrapolation could introduce a bias in the light curves. For the flaring states the hardest spectral index reported is $\Gamma = 2.21 \pm 0.18$ \citep{albert:2008:magic:m87} and, therefore, the systematic error introduced when using $\Gamma = 2.3$ is completely negligible compared to the typical statistical error of 10-20\% of the flux during such states. For the quiescent state spectral indices up to $\Gamma = 2.60 \pm 0.30$ have been reported \citep{albert:2008:magic:m87}. This could, in principle, create a bias for the \He data set, where the flux is extrapolated down to 350\,GeV from energies above 500\,GeV. For example, for spectral indices of $\Gamma_1 = 2.3$ and $\Gamma_2 = 2.6$ the flux extrapolated down to 350\,GeV from 700\,GeV would differ by a factor $(350\,\text{GeV} / 700\,\text{GeV})^{\Gamma_2 - \Gamma_1} \sim0.8$, which would imply errors of order $\sim$20\,\%. Given that the typical flux error for the nightly averaged flux bins in the quiescent state is of order $\sim$100\,\% (i.e. the individual nightly flux points are not significant detections) the systematic uncertainty introduced by using a single photon index for extrapolation can also safely be neglected.

Given recent results on the Crab Nebula (the reference source for ground-based VHE instruments) indicating only small systematic offset between the energy scale of different instruments \citep{meyer:2010a}%
\footnote{Energy scale scaling factors relative to the \FermiLAT energy scale of $0.961 \pm 0.004$ for \He and $1.03 \pm 0.01$ for \Ma are derived.}
and the general good agreement of the flux measurements during quasi simultaneous observations, the systematic error between the different instruments is estimated to be small compared to the statistical error of individual measurements of the 2010 campaign.

Additional archival data from \citet{aharonian:2006:hess:m87:science}, \citet{acciari:2008a:veritas:m87}, \citet{albert:2008:magic:m87}, \citet{acciari:2010:veritas:m87:2008-2009}, and \cite{aleksic:2011:magic:m87:mono} are also shown in the light curves (Fig.~\ref{Fig:MWL-longterm-lightcurve} and Fig.~\ref{Fig:VHELightcurves2005-10}).

In the following, the main characteristics of the VHE observatories involved in the \M campaign are reviewed, and details on the corresponding data-sets are presented.

\paragraph{H.E.S.S.}
The High Energy Stereoscopic System (H.E.S.S.)\footnote{http://www.mpi-hd.mpg.de/hfm/HESS/} is a system of four large (13\,m mirror diameter) imaging atmospheric Cherenkov telescopes (IACTs) for the detection of VHE $\gamma$-rays, located in the southern hemisphere in Namibia ($23^{\circ}16'\,$S, $16^{\circ}30'\,$W; 1800\,m above sea level). It has been in operation since 2002, with the full array completed in 2004 \citep{hinton:2004a:hess:status}. H.E.S.S. observed \M for 10.7\,h in 2010 (dead-time corrected) yielding a total detection significance of 9.7\,standard deviations (s.d.; following \citealt{li:1983a}) using a standard Hillas-type analysis and cuts from \citealt{aharonian:2006:hess:crab}  (software version \texttt{hap-10-06}). \M culminates at $\sim$35\degree zenith angle at the H.E.S.S. site resulting in an energy threshold of $E_{\rm thr} > 500$\,GeV for the analysis and data-set. The integral flux is extrapolated down to 350\,GeV using the average energy spectrum (see above). In addition, results from a re-analysis of the 2004 and 2006 \He data in nightly flux bins, utilizing the same analysis as discussed above, are presented in the light curves. Cross check on the results with data from an independent calibration chain and utilizing different analysis have been performed and good agreement is found.

\paragraph{MAGIC}
The Major Atmospheric Gamma-ray Imaging Cherenkov (MAGIC)\footnote{http://wwwmagic.mppmu.mpg.de/} telescope system consists of two 17\,m diameter IACTs located at the Roque de los Muchachos Observatory, on the Canary
Island of La Palma ($28^{\circ}46'\,$N, $17^{\circ}53'\,$W; 2200\,m above sea level).
Since 2005 \M has been regularly observed by MAGIC with a single telescope
\citep{albert:2008:magic:crab, aleksic:2011:magic:m87:mono}. In the Autumn of 2009 the system became stereoscopic,
with the commissioning of a second telescope and the stereo trigger, resulting in an almost doubling
of its sensitivity \citep{colin:2009a, aleksic:2011:magic:stereo:performance}. In 2010, \M observations were conducted,
for the first time, in stereoscopic mode. 20\,h of good quality data were taken between January
and June with zenith angles ranging from 16 to 35 degrees. Analysis of these data with the standard
MAGIC software \citep{moralejo:2009a:magic:mars} resulted in a 10\,s.d.\ detection above 200\,GeV.

\paragraph{VERITAS}
The Very Energetic Radiation Imaging Telescope Array System (VERITAS)\footnote{http://veritas.sao.arizona.edu/}
consists of four $12 \, \rm{m}$ diameter IACTs and is located at the base camp of the Fred Lawrence
Whipple Observatory in southern Arizona
 ($31^{\circ}40'\,$N, $110^{\circ}57'\,$W; 1280\,m above sea level). More details about VERITAS, the data calibration, and the analysis techniques can be found in \citet{acciari:2008a:veritas:m87}. VERITAS observed \M in 2010 for $48.2 \,\rm{h}$ (after quality cuts) resulting in a detection of 26\,s.d. The zenith angles of the observations ranged from 19 to 40\degree, with  a few nights with zenith angles up to 60\degree during the flares (April 9-11th). The energy threshold of the analysis for the mean zenith angle of the observations of 25\degree is 250\,GeV. The VERITAS data covered the whole 2010 flare period; a detailed study of the VHE spectral evolution indicating a spectral change with flux will be published in a parallel paper \citep{aliu:2011a:veritas:m87:2010}.

 \subsection{High energy (HE)}
 
\paragraph{\FermiLAT}
The Large Area Telescope (LAT) onboard the \Fermi satellite is a pair-conversion telescope that covers
the energy range from 20 MeV to more than 300\,GeV (HE) \citep{atwood:2009:fermi:lat:technical}.  The LAT
instrument features a per-photon angular resolution of $\theta_{\rm
68\%}=0.8\degree$ at $1$\,GeV and a large field-of-view of $2.4$\,sr.  The primary
mode of operation is an all-sky survey mode, where the full sky is covered
approximately every three hours.  The LAT data for this analysis consists of two
years of nominal all-sky survey data between the energy range $100$\,MeV and
$300$\,GeV, and spanning the mission elapsed time (MET) 239557417 to 302630530
(August 4, 2008 through August 4, 2010).  Event selections include ``diffuse''
class events recommended for point source analysis, a rocking angle cut of
$<52\degree$, and a zenith angle cut of $<100\degree$ in order to avoid contamination
from the Earth's limb.  A $2$\,ks window beginning at MET 259459364 was also
removed in order to avoid contamination from GRB\,090323, which occurred
nearby.  All LAT analysis was performed using instrument response functions
(IRFs) \texttt{P6\_V11\_DIFFUSE} and science tools \texttt{v9r20p0}, along with
the recommended\footnote{\url{http://fermi.gsfc.nasa.gov/ssc/data/access/lat/BackgroundModels.html}} Galactic diffuse \texttt{gll\_iem\_v02\_P6\_V11\_DIFFUSE.fit} and corresponding isotropic
spectral template \texttt{isotropic\_iem\_v02\_P6\_V11\_DIFFUSE.txt}.

An analysis of the LAT spectrum over the two-year period was performed using a
binned likelihood analysis \citep{mattox:1996a} selecting all events that fell
within a $20\degree \times 20\degree$ square region of interest (ROI) centered at the
\M radio position of the core. All point sources from an internal two-year preliminary
catalog that fell within $15\degree$ of the source were included in the fit.  All
sources within the square ROI were modeled with a power law spectrum with their normalization
and index as free parameters, while those that fell outside of the ROI were
fixed to their catalog values. The \M spectrum was modeled as a power law
with photon index and normalization parameters left free and using the radio
position \citep{fey:2004b} as the source location.  A point source was detected
with a test statistic (TS; \citealt{mattox:1996a}) of 301, representing a detection
of $\sqrt{301} \simeq 17$\,s.d.\  From the resulting fit, the photon index and
flux ($> 100$\,MeV) were found to be $2.16 \pm 0.07$ and $(2.66 \pm 0.36) \times
10^{-8}$\,ph\,cm$^{-2}$\,s$^{-1}$, respectively. In order to test for curvature, the
spectrum was also fit to a log parabola, where the TS of the overall fit was found
to improve by only 0.89, which does not represent a statistically significant 
improvement over the single power law. The largest systematic errors
can be attributed to uncertainties in the modeling of the diffuse background
emission. These were estimated by repeating the analysis with both binned and
unbinned \texttt{gtlike} using a refined version of the current diffuse
background model that is under development by the LAT collaboration.
Systematic errors on the index and flux were thus found to be $(+0.05 / -0.01)$
and $(+0.40 / -0.13) \times 10^{-8}$\,ph\,cm$^{-2}$\,s$^{-1}$, respectively.
Comparing results from this analysis for the last 14 month with the initial $10$-month spectrum reported in \citet{abdo:2009:fermi:lat:m87}, no evidence of variability in the flux above $100$\,MeV was found. Comparing
the flux above 1\,GeV between these two epochs, however,  a marginal
indication of a rise in the flux of the latter epoch at a significance of
2\,s.d.\ is found. Systematic uncertainties on the LAT flux due to the instrument
response are estimated at values of 10\,\%, 5\,\%, and 20\,\% above and
below their nominal values at $\log (\rm E/MeV)=2, 2.75,$ and $4$,
respectively \citep{atwood:2009:fermi:lat:technical}.

The LAT light curve ($> 1$\,GeV) was constructed using the events
that fell within a $10\degree$ circular ROI centered at the \M radio position.
Fluxes were derived for bins with a detection significance exceeding 3\,s.d., otherwise upper limits were calculated.
To generate the light curve, events were grouped into 56-day time bins, and a
separate likelihood analysis using \texttt{gtlike} was performed over each of
the bins.  All point sources from the two-year fit were included in the model
over each interval.  Sources that fell within the $10\degree$ ROI were fit with
their normalization parameters free, while the photon index of each source was
fixed to the best-fit value obtained from the full two-year analysis.  Both the
index and normalization parameters of \M were left free, except in the case of
upper limit calculations, in which case the photon index was fixed to the
nominal two-year average of 2.16.  In order to avoid modeling sources with a
negative TS value, an initial fit over each interval was performed, and all
sources found to have a TS$<1$ were subsequently removed from the fit.
Following the method for variability detection outlined in \citet{abdo:2010:fermi:1fgl},
the weighted average flux was first calculated with a resulting value of
$(1.62\pm0.18) \times 10^{-9}$\,ph\,cm$^{-2}$\,s$^{-1}$.  A $\chi^{2}$ analysis
was then performed by comparing the best-fit values of all points to the
weighted average, and the resulting probability $P(\chi^2 \ge \chi^2_{\rm obs})$
was found to be 0.027, which represents a significance of 2.2\,s.d.\ for the source to be variable and falls
slightly below the threshold for variability defined in \citet{abdo:2010:fermi:1fgl}.

\subsection{X-ray}\label{Sec:DataXRay}
 
\paragraph{Chandra}
 
X-ray data have been taken with the  Advanced CCD Imaging Spectrometer (ACIS) on board  the Chandra satellite. For details of the \Chandra data reduction procedures, see \citet{harris:2003a}, \citet{perlman:2003a}, and \citet{harris:2009a}.
In brief, a 1/8th segment of the back illuminated S3 chip of the
ACIS detector aboard \Chandra is used.  This results in a frame time
of 0.4s with 90\,\% efficiency.  Although this setup was essentially
free of pileup when \citet{wilson:2002a} tested various options during 2000
July, with the advent of the ever increasing brightness of HST-1,
pileup \citep{davis:2001a} became a major problem so the measure of intensity was switched to a
detector based unit: keV/s.  This approach uses the
event 1 file with no grade filtering (so as to recover all events
affected by 'grade migration') and energies from 0.2
to 17 keV are integrated so as to recover all the energy of the piled events.  Other
uncertainties for piled events come from the on-board filtering, the
'eat-thy-neighbor' effect, and second order effects such as release
of trapped charge (see Appendix~A of \citealt{harris:2009a}).
Although a small circular aperture was used for flux-map photometry in
 \citet{harris:2003a}, the basic analysis for this paper adopts the rectangular
regions used in \citet{harris:2006a} so as to encompass more of the point spread
function (PSF). 
All events within each rectangle are weighted by their energy and the sum
of these energies, when divided by the exposure times, gives the final
keV/s value used in the light curve.  Uncertainties are strictly
statistical, based on the number of counts measured ($\sqrt{N}/N$) and
typically range from 1\% to 5\%.

\subsection{Optical}\label{Sec:DataOpticalLT}

\paragraph{Hubble Space Telescope (HST)}
Near Ultraviolet HST observations were obtained with 
the Space Telescope Imaging Spectrograph (STIS) and the Advanced Camera for 
Surveys High Resolution Camera (ACS/HRC). STIS imaging was taken  with the  
F25QTZ filter that has its maximum throughput wavelength at $2364.8\,\mbox{\AA}$ 
\citep{kim:2003a}. All observations after August 2004 were taken with  
the ACS/HRC using two filters: F220W and F250W which have their maximum throughput 
wavelength at $2255.5\,\mbox{\AA}$ and  $2715.9\,\mbox{\AA}$, respectively 
\citep{mack:2003a}. All science ready files were retrieved from the STScI public
archive and processed through the PYRAF task MULTIDRIZZEL
\citep{fruchter:2009a}. Both \HST detectors have a pixel scale of $\sim0.024\arcsec$
per pixel and a resolution that enables one to clearly separate the nucleus 
 and the innermost components of the jet, i.e. HST-1. The details 
of these observations have been presented in \cite{perlman:2003a}, \citet{madrid:2009a}, and
\citet{perlman:2011a}.

Optical \HST polarimetry observations were obtained with the ACS/HRC and the Wide-field Planetary Camera 2 (WFPC2), using the F606W filter that has its maximum throughput at roughly $\sim$6000\,$\AA$ and a nearly flat throughput curve from 4800 to 7000\,\AA.  The single WFPC2 observation (pixel scale $\sim 0.1$\,arcsec/pixel) was obtained when the ACS was not operational, and used the POLQ polarizing filter, while the ACS/HRC observations used the POLVIS polarizing filter.  For both of these polarizers, the F606W filter gives nearly optimal transmission of parallel polarized light as well as rejection of perpendicularly polarized light.  Polarimetry observations were obtained at 18 epochs between December 2002 and November 2007, with 14 of the 18 observations concentrated between November 2004 and December 2006 on the same schedule as the ultraviolet photometry.  As with the UV observations, all polarimetry was obtained from the \HST archive and re-calibrated using updated flat field files and image distortion correction (IDC) tables \citep{mobasher:2002a,pavlovsky:2002a}.  {\sc multidrizzle} was used to combine and cosmic-ray reject the images, which were aligned using Tweakshifts \citep{fruchter:2009a}, and CTE (Charge Transfer Efficiency) corrections were computed using data found in the ACS instrument handbook \citep{boffi:2007a}.  Once this was done, the images in each polarizer were combined according to recipes in the ACS and WFPC2 instrument handbooks, respectively \citep{boffi:2007a,biretta:1997a}. Before performing photometry and polarimetry, galaxy emission was subtracted from the images.  This was done using a model computed in the Stokes I image using the IRAF tasks ELLIPSE and BMODEL. Error bars for the polarimetry are typically 2-3\% for high signal-to-noise data \citep[further details can be found in][]{perlman:2011a}.

\paragraph{Liverpool Telescope (LT)}
Hybrid R+V-band  optical polarimetry data (460-720\,nm at FWHM) were taken
with the 2-meter Liverpool Telescope (LT)\footnote{http://telescope.livjm.ac.uk/}, located on
La Palma, using the newly commissioned RINGO2 fast-readout imaging
polarimeter \citep{steele:2010a}. The polarimeter uses a rotating
polaroid with frequency of approximately 1\,Hz, during the cycle of which 8
exposures of the source are obtained. These exposures are synchronized
with the phase of the polaroid and following the analysis method of \citet{clarke:2002a}
 allows determination of the degree and angle of polarization.

The LT observations were taken during and shortly after the observed VHE
high-state, MJD 55295-402, with three measurements taken
contemporaneously to the time of the VHE flare, MJD 55295-97.
Further data on \M have been taken since then and will be presented elsewhere.
Total integration times of typically 100\,s were used in the
observations, corresponding to an achieved polarimetric accuracy of about
1\,\% for the brightness of the source. The field of view of the instrument
is of $4 \times 4$\,arcmin with a pixel scale of 0.45\,arcsec/pixel, which at the
distance of \M corresponds to a linear scale of $\sim$40\,pc/pixel,
equivalent to approximately half the distance between the core and the
innermost jet component, the knot HST-1. Given the typical seeing during the observations of $1.0\,\arcsec$ to $1.7\,\arcsec$ (FWHM) the two components, core and HST-1, cannot be resolved individually. Therefore, an aperture radius for the integration of the signal of 2.7\,arcsec was used, so that both the nucleus and HST-1 were included. The outer jet is
nevertheless well resolved and independent light-curves were produced with
the same aperture radius used for the core but now centered at knot-A,
located 12.3\,arcsec downstream from the nucleus, revealing a steady polarization
aligned with the jet direction.

The greatest source of error in the determination of the polarization
levels is a systematic effect caused by contamination by the bright host
galaxy, which extends out to several kpc and dilutes the polarization
signal from the nucleus.  The strength of the contamination is estimated by adding a Sersic profile
with $n=4$ to a frame built by summing data from all polaroid angles. This
shows the well known flattening in the central $\sim10$\,arcsec (e.g. \citealt{kormendy:2009a}).
Fitting the data from 9 to 2.7\,arcsec reveals an excess
within the central 2.7\,arcsec due to the core plus HST-1 of $\sim10\pm5$\,\%. 
The photometric flux of the background galaxy is of order 90\,\% of the
total flux within the aperture used for extracting the polarization
measurements.  The measured fractional polarization was therefore multiplied by a
factor of 10 and the flux measurement divided by a factor 10 in order to correct for this background light, and allow
comparison with the high resolution \HST measurements of the core + HST-1.

Light-curves for the polarization position angle (or electric vector position angle, EVPA, defined as
$0.5 \times \arctan(u/q)$, where $u$ and $q$ are the linear Stokes parameters) were
obtained for all the epochs of observation, and are presented in Fig.~\ref{Fig:OpticalPolarimetry},
along with the other polarization quantities. Due to host-galaxy
contamination,the polarization position angle for the nucleus and HST-1 combined are measured to
an accuracy of 25\degree, considerably degraded in comparison to the 10\degree resolution achieved for the outer jet where the
host galaxy is fainter.

 \subsection{Radio}\label{Sec:DataRadio}

\begin{deluxetable}{lcc}
\tablecaption{Angular resolution for different radio observations.}
\tablehead{
\colhead{Instrument} &
\colhead{Wavelength} &
\colhead{Resolution}
}
\tablecomments{The beam sizes are given as full width at half maximum (FWHM). In case different resolutions have been used in the analysis the highest one is given in the table.}
\startdata
VLBA & 43\,GHz & $(0.21 \times 0.43)$\,mas \\
VLBA (MOJAVE) & 15\,GHz &(0.6 $\times$ 1.3)\,mas \\
VLBA & 2.3\,GHz & (7.5 $\times$ 3.9)\,mas \\
VLBA & 1.7\,GHz  & (8.0 $\times$3.4)\,mas \\
EVN & 5\,GHz & $(1.0\times2.0$)\,mas \\
VLA & 15\,GHz & $\sim 0.13\arcsec \times 0.12\arcsec$ \\
VLA & 22\,GHz & $\sim 0.10\arcsec \times 0.09\arcsec$
\enddata
\label{Table: CoreAngularResolution}
\end{deluxetable}
 
Radio interferometers enable one to observe the jet of \M with a large variety of angular resolutions and sensitivities, depending on the array size and observing frequency. In Tab.~\ref{Table: CoreAngularResolution} the highest angular resolution achieved for each instrument contributing data to this paper is shown. Overall, they span an angular scale range from as large as $\sim 0.1\arcsec$ (from the VLA at 22 GHz) down to a fraction of a milli-arcsecond (e.g. $\sim 0.2$ mas with the VLBA at 43\,GHz, see Table 1). The longest baseline and highest frequency observations provide the most valuable information about the compact, flat spectrum core; conversely, instruments with shorter baselines (such as the VLA) or lower observing frequency (like the EVN) are most valuable for the fainter and more extended HST-1 feature.
 
\paragraph{VLBA 43\,GHz}

The 43 GHz VLBA \citep{napier:1994a} data from 2006 through 2008 were
collected as part of an effort to study the dynamics of the inner jet
near the launch region \citep{walker:2008a}.  The 2009 and 2010
observations were part of a project to find and study another
VHE/radio flare like that seen in 2008 \citep{acciari:2009b:m87joinedcampaign:science}.
The observations were made on the 10-antenna VLBA using a total
bandwidth of 64\,MHz.  The data were reduced in AIPS following the
usual procedures for VLBI data reduction including correction for
instrumental offsets using the autocorrelations, bandpass calibration
based on strong calibrator observations, and correction for
atmospheric opacity based on the system temperature data. The a
priori amplitude calibration depended on the gains provided by VLBA
operations, which are based on results from regular single-dish
pointing observations of Jupiter, Saturn, and Venus averaged over many
months.  The images are based on data that are both amplitude and phase
self-calibrated.  The flux scale for each epoch was set by normalizing
the self-calibration gain adjustments to the a priori gains for
observations above 30\degree elevation on those antennas with good
weather and instrumental conditions for that epoch.  The flux
densities typically accurate to within about 5\%.

Three VLBA flux densities are provided.  The first is the peak
brightness on the core in an image made with a $0.21 \times 0.43$\,mas
 beam.  The second is the integrated flux density
within 1.2\,mas of the core and represents the total emission from a
region within a projected distance of 0.1\,pc from the presumed position
of the black hole at the radio core
(0.1\,pc = 340\,R$_{\rm S}$ for a $3.0 \times 10^{9} \, \rm{M}_\odot$ black hole).
The third is the integrated flux density from the jet
over the region 1.2 to 5.3\,mas from the core.

\paragraph{VLBA 15\,GHz (MOJAVE)}
VLBA 15\,GHz observations from the MOJAVE
program \citep{lister:2009a} have been analyzed to obtain core fluxes over 16 epochs from
2001.0 - 2011.0. The calibrated ($u, v$) data were
retrieved\footnote{http://www.physics.purdue.edu/MOJAVE/} and re-imaged
uniformly with the final maps restored using a 0.6\,mas $\times$ 1.3\,mas
beam (position angle $= -11$\degree) following \citet{abdo:2009:fermi:lat:m87}. The peak
core fluxes measured from the resultant maps span typical values of
$\sim$1.0--1.2\,Jy/beam (Fig.~\ref{Fig:MWL-longterm-lightcurve}) with two notable peaks of $\sim$1.5\,
Jy/beam recorded in early-2008 and mid-2009.

\paragraph{VLBA 2.3\,GHz}
Observations at 2.3\,GHz were made on 8 and 18 April 2010 using 10 VLBA stations as a part of experiments BH163.  
Each session has a total on-source time of $\sim$15~minutes with the total bandwidth being 64\,MHz. 
In order to obtain better $u, v$ coverage,  the short scan blocks ($\sim$ 2 minutes per block) were distributed uniformly at several hour angles. The initial data calibration was performed in AIPS based on the standard VLBI data reduction procedures (the AIPS cookbook\footnote{http://www.aips.nrao.edu/cook.html}). The amplitude calibration with the opacity corrections was applied 
using the measured system noise temperatures and the elevation-gain curves of each antenna.The data were fringe-fitted and then averaged at short intervals (every 5\,seconds and 1\,MHz in time and frequency domains) 
before the imaging process in order to avoid the smearing effects due to the time and bandwidth averaging at the HST-1 region. 
The images were made in DIFMAP software \citep{shepherd:1997a} with the iterative phase/amplitude self-calibration processes.  The off-source rms noises in the resultant images are 0.4-0.5\,mJy/beam. 
The peak flux densities are provided for the core region with the synthesized beam of 7.5 $\times$ 3.9\,mas at $-5^{\circ}$, and the integrated flux densities are provided for the HST-1 region. The errors in flux densities are assumed to be 5\% based on the typical VLBA calibration accuracy at this frequency. 

\paragraph{VLBA 1.7\,GHz}
19 epochs of VLBA 1.7\,GHz observations from
$\sim$2005.0 - 2008.0 have been obtained in an effort to monitor the evolution of the HST-1
knot following its brightening at X-ray, optical, and radio frequencies
\citep{harris:2009a}.  The results of the first 9 of these observations
(programs BH126, BH135) were presented in \citet{cheung:2007a}, where
apparent superluminal radio features in the HST-1 complex were discovered. For
the additional 10 observations (programs BC167, BH151), the identical
calibration and imaging procedure was followed for this analysis. From
maps restored with the same uniformly weighted beam (8.0\,mas $\times$
3.4\,mas elongated north-south), we measured the peak core brightness
(5$\%$ errors are assumed). Integrated flux densities (10$\%$ errors are
assumed) for the HST-1 were measured from naturally weighted images
(11.5\,mas $\times$ 5.5\,mas) using a 80\,mas $\times$ 50\,mas box that
covers the entire radio complex resolved in the VLBA images (cf., \citealt{cheung:2007a}).

\paragraph{EVN}
\M was observed with the European VLBI Network (EVN) at 5\,GHz seven times between November 2009 and June 2010 as part of a project aimed at correlating the radio and high energy behavior of the source. Given the interesting episodes of activity of the source during the campaign, a few observations were scheduled as ToO and do not have the same array configuration as the other ones. In general, 6 to 11 telescopes participated in the observations, with baselines ranging between less than 100\,km (as provided by MERLIN stations) and 7\,000-9\,000\,km (as provided by the Arecibo and Shanghai stations, respectively). The observations were carried out at 5.013\,GHz, divided in 8 sub-bands separated by 16\,MHz each, for an aggregate bit rate of 1\,Gbps. The data were correlated in real time at JIVE using the so-called e-VLBI technique, which provides a fast turn-around of the results \citep[as was the case near the Feb.\ 2010 event; see][]{giroletti:2010a}. Automated data flagging and initial amplitude and phase calibration were also carried out at JIVE using dedicated pipeline scripts. Data were finally averaged in frequency within each IF, but individual IFs were kept separate to avoid bandwidth smearing. Similarly, the data were time-averaged only to 8\,s, in order to avoid time smearing. Final images were produced in DIFMAP after several cycles of phase and amplitude self-calibration. Various weighting schemes were applied to the data to improve resolution in the core region and enhance the fainter emission in the HST-1 region. For the core, uniform weights were used obtaining a typical HPBW of $1.0\times2.0$\,mas (in PA\,$-20^\circ$) and providing the peak brightness. For HST-1, natural weights were used resulting in a $7.5\times8.5$\,mas HPBW; a resolved structure is clearly detected and integrated flux densities for the full region are given. For additional details, please see \citet{giroletti:2011a}.

\paragraph{VLA}
Data from the VLA archive for 10 epochs have been used, selecting observations performed in A-array at 15 and/or 22\,GHz. The typical angular resolution of the VLA is $\sim 0.13\arcsec \times 0.12\arcsec$ and $\sim 0.10\arcsec \times 0.09\arcsec$ at the two frequencies, respectively, which permits one to resolve the core and the HST-1 feature. Data reduction was carried out in AIPS in the standard manner: the flux density scale was tied to the main gain calibrator 3C\,286 (which was used in all the observations) using SETJY with the available model. Phases were calibrated using the compact source 1224+035, obtaining good solutions. A few final iterations of the phase and amplitude self-calibration have been carried out to improve the image quality. Integrated flux densities for the core and HST-1 have been derived using JMFIT.

Additional 15\,GHz VLA data for HST-1 from \citet{harris:2009a} are also shown in the light curve.


\section{The 2010 VHE campaign}\label{Sec:2010VHECampaign}

\begin{figure*}
\centering
\includegraphics[width=0.8\textwidth]{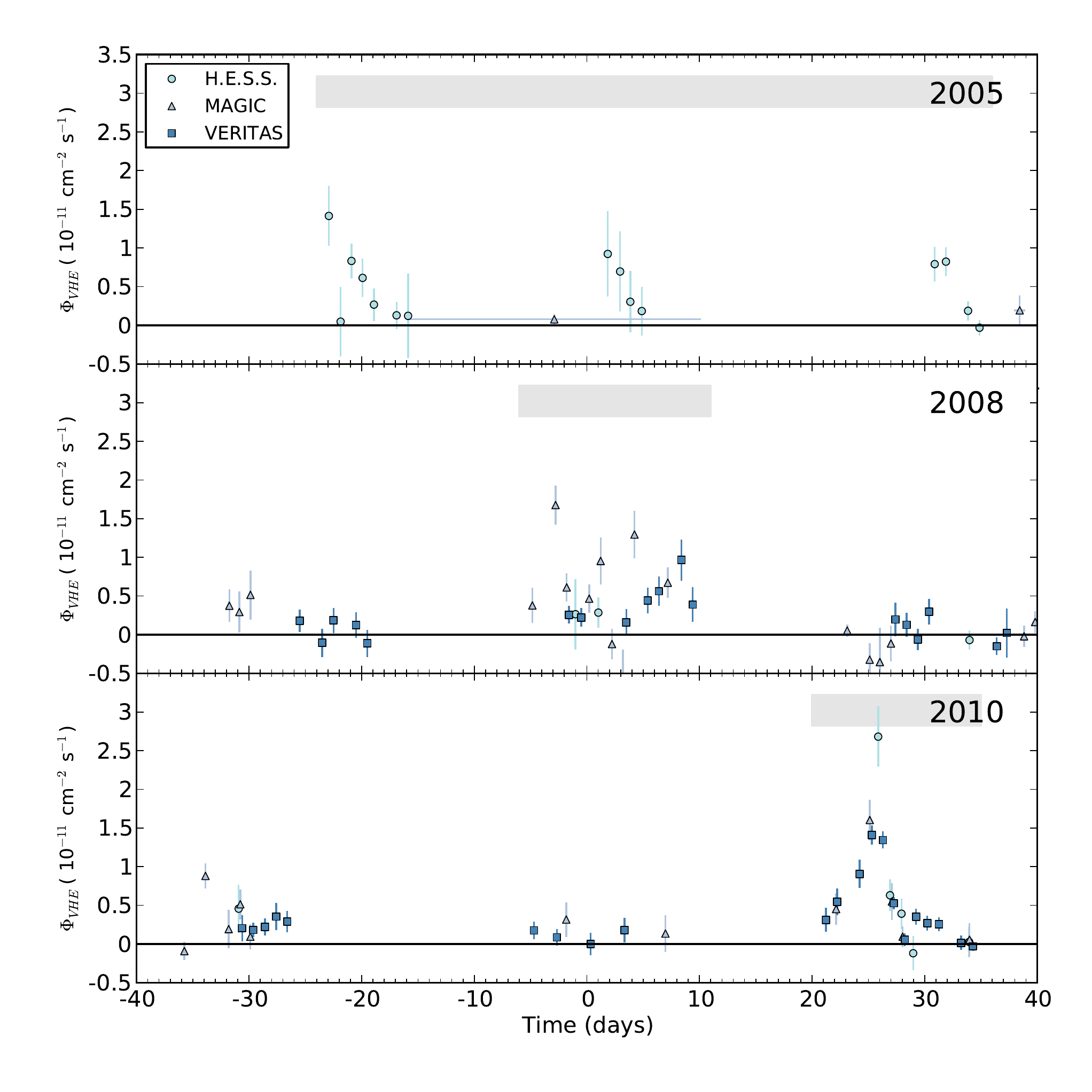}
\caption{VHE light curve of M\,87 of the flaring episodes in 2005 (top), 2008 (middle), and 2010 (bottom). Integral fluxes are given above an energy of 350\,GeV. The lengths of the gray bars correspond to the length of the gray shaded areas in Fig.~\ref{Fig:MWL-longterm-lightcurve}. A time of 0\,days corresponds to MJD\,53460, MJD\,54500, and  MJD\,55270 for 2005, 2008, and 2010, respectively. Flux error bars denote the 1\,s.d.\ statistical error. Horizontal error bars denote the time span the flux has been averaged over. Note that in the case of time spans longer than one night the coverage is not continuous.}
\label{Fig:VHELightcurves2005-10}
\end{figure*}
 
\begin{figure}
\centering
\includegraphics[width=0.5\textwidth]{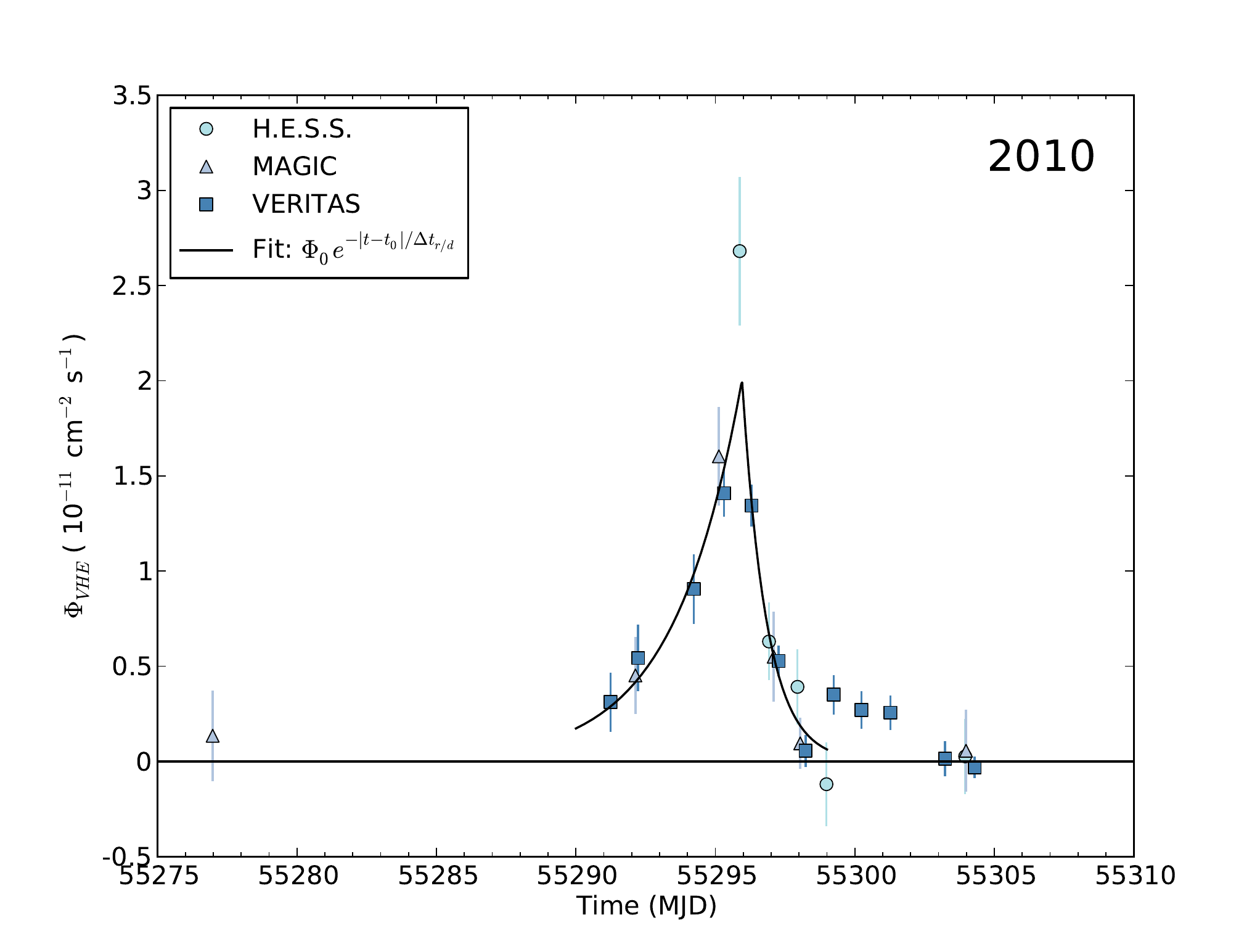}
\caption{VHE light curve of M\,87 zoomed on the 2010 flare. Also shown are the results of the fit of an exponential function to the data. Error bars denote the 1\,s.d.\ statistical error. The results of the fit to the data are summarized in Tab.~\ref{Table:FitFlareExp}. Further details can be found in the text.}
\label{Fig:VHELightcurve2010}
\end{figure}
 
\begin{deluxetable}{lcc}
\tablecaption{Parameters of the fit to the 2010 VHE flare data}
\tablehead{
\colhead{Parameter} &
\colhead{Value} &
\colhead{Unit}
}
\tablecomments{The fit results in a $\chi^2/\text{d.o.f.} = 10.02/11$ with chance probability $P = 0.53$. The parameters are defined in the text.}
\startdata
Fit range & 55290 - 55299 & MJD  \\ \tableline
$\tau_\text{d}^{\text{rise}}$ &  $1.69 \pm 0.30 $ & d \\ 
$\tau_\text{d}^{\text{decay}}$ &  $ 0.611 \pm 0.080 $ & d \\
$t_0$ & $55295.954 \pm 0.094$ & MJD \\
$\Phi_0$ & $(2.01 \pm 0.15) \times 10^{-11}$ & cm$^{-2}$ s$^{-1}$
\enddata
\label{Table:FitFlareExp}
\end{deluxetable}
 
\paragraph{The 2010 flare}
During the 2010 VHE monitoring campaign two episodes of increased VHE activity have been reported \citep{mariotti:2010:magic:m87:atel,ong:2010a:m87:veritas:magic:flare:atel}: The first episode took place in Feb. 2010 where a single night of increased activity was detected by \Ma (Fig.~\ref{Fig:VHELightcurves2005-10} bottom panel around $-35$\,d; detection significance $>5$\,s.d.). Follow-up observations by \He and \Ve did not reveal further activity at VHE. The second episode took place in Apr. 2010 and showed a pronounced VHE flare detected by several instruments triggering further MWL observations. In the following, the discussion will concentrate on this second flaring episode.

The VHE activity of this second flaring episode is concentrated in a single observation period between MJD 55290 and MJD 55305 ($\sim$15\,days; see Fig.~\ref{Fig:VHELightcurves2005-10} bottom panel \& Fig.~\ref{Fig:VHELightcurve2010}). This time-period is exceptionally well covered with 21 pointings by different VHE instruments, resulting in an observation almost every night. It should be noted that during nights with (quasi) simultaneous observations by different instruments, the measured fluxes are found to be in very good agreement.

The detected flare displays a smooth rise and decay in flux with a peak around MJD 55296 (April 9-10, 2010). A peak flux of $\sim 2.5 \times 10^{-11}$\,cm$^{-2}$s$^{-1}$ is reached, which is about a factor 10 above the quiescent flux level of the source. The data on the rising part of the flare indicates a steady rise. On the decaying side the situation is more complex: two nights after the detection of the maximum flux (MJD 55298) all three instruments measured a low flux compatible with zero, while in the following 3 nights (MJD 55299/55300/55301) a higher flux is detected by \Ve.

To derive the timescales of the flare the VHE light curve from MJD 55290 to MJD 55299 is fitted with a two-sided exponential function
\begin{equation}
\Phi =  \Phi_0 \times e^{-|t - t_0|/\Delta \tau} \text{ with }
\begin{cases}
\Delta \tau = \Delta \tau^\text{rise} \text{ for } t < t_0 \\
\Delta \tau = \Delta \tau^\text{decay}  \text{ for } t  \geqslant t_0
\end{cases}
\end{equation}
resulting in an excellent description of the data  with a $\chi^2/\text{d.o.f.} = 10.02/11$ and a corresponding chance probability of $P = 0.53$. The resulting fit parameters are summarized in Tab.~\ref{Table:FitFlareExp}.
Very similar results are obtained when (i) fitting a longer time span (e.g. MJD 55275 - 55305) or when (ii) adding a constant offset as a free parameter to account for a possible baseline flux, though the chance probabilities of these fits are slightly worse then for the original fit described above.

Flux doubling times of $\tau_\text{d} = ln(2) \times \Delta \tau$ of $\tau_\text{d}^{\text{rise}} = (1.69 \pm 0.30)$\,days for the rising and $\tau_\text{d}^{\text{decay}} = (0.611 \pm 0.080)$\,days for the decaying part of the flare are derived, which signifcantly differ by a factor $2.77 \pm 0.62$. The variability timescales derived for the 2010 VHE flare are the most precisely measured VHE variability timescales determined for \M to date. Previously detected VHE flares only allowed for rough estimates due to the variability pattern, sampling, and statistics.

\paragraph{Comparison with the 2005 and 2008 flares}
The VHE light curve from 2005, 2008, and 2010 around the flaring episodes is displayed in Fig.~\ref{Fig:VHELightcurves2005-10}. In all three flares, similar flux levels are reached. The apparent timescales of the order of days are also very similar. The time period over which activity is detected, is comparable in 2008 and 2010 but is slightly longer in 2005, though the gaps in the sampling make any definitive statement difficult. During the 2010 activity period a pronounced flare is detected which is well described by an exponential behavior (see previous paragraph). While the 2005 flare is compatible with such behavior (the sampling is considerably worse than in 2008 and 2010), the 2008 activity state seems more erratic with several maxima over a similar time period as in 2010. Taking the best fit function to the 2010 flare as a template, large parts of the 2008 flaring activity are not compatible with the 2010 behavior (see the Appendix for details). 

\paragraph{VHE flare duty cycle}
Given the typical length of a VHE flare of order a day, the nightly VHE $\gamma$-ray flux measurements presented in this paper can be used to estimate the duty cycle of \M for VHE flares. For example, following the approach presented in \citet{jorstad:2001a}\footnote{Calculating the number of observations with a flux a factor 1.5 above the error weighted mean (for this data set $\sim1.8 \times 10^{-12}$\,cm$^{-2}$s$^{-1}$).}, one derives a duty cycle of $\sim$28\%. However, due to the observing strategy followed for \M (observations have been intensified after the detection of a high state), the data-set is biased. The derived duty cycle is, therefore, likely overestimated and should be considered an upper limit. To enable a more unbiased view and to give an estimate of the possible uncertainties the duty cycle for a range of threshold fluxes is presented, calculated from the number of data points above the threshold flux divided by the total number. For a threshold flux of $\phi_\text{thresh} = (0.5 / 0.8 / 1) \times 10^{-11}$\,cm$^{-2}$s$^{-1}$ the respective duty cycles are $\sim$14\,\% / 7\,\% / 4\,\%.


\section{Optical polarimetry}\label{Sec:OpticalPolarimetry}

\begin{figure*}
\centering
\includegraphics[width=0.8\textwidth]{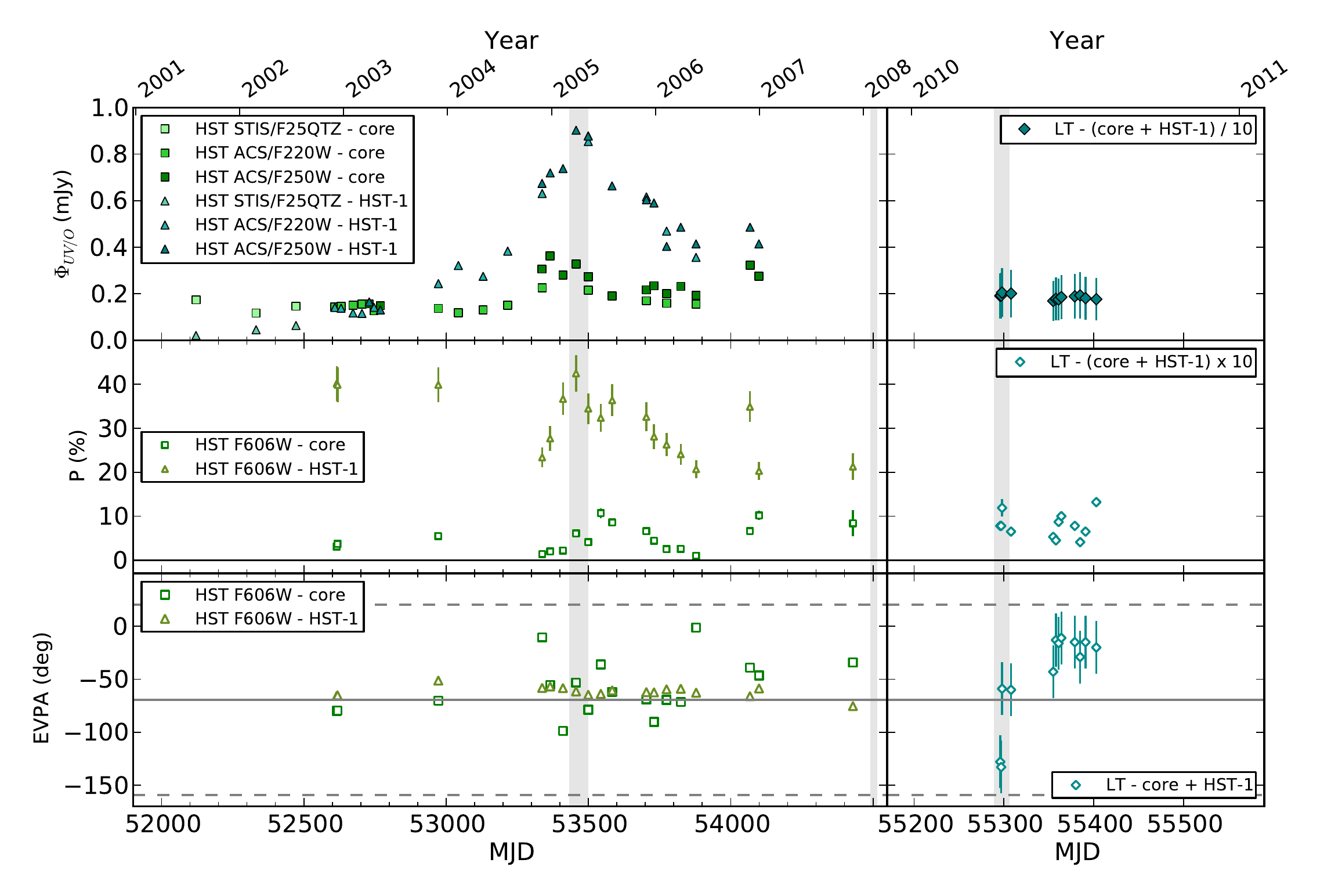}
\caption{Optical polarization for different components of the \M jet as measured by \HST and LT. The LT polarization measurement of the core + HST-1 complex is scaled up by a factor of 10 to compensate for the host galaxy emission in the measurements (see Sect.~\ref{Sec:DataOpticalLT} for details). The horizontal gray solid line in the lower panel marks the jet direction ($-69.5$\degree) while the horizontal gray dashed lines are located at $-69.5\degree \pm 90$\degree (180\degree distance).}
\label{Fig:OpticalPolarimetry}
\end{figure*}

Radio and optical emission from regions within the \M jet show a high degree of polarization \citep{perlman:1999a}, which is indicative of non-thermal emission processes in highly ordered magnetic fields (synchrotron radiation). In general, polarization and especially changes in the polarization can be utilized to investigate the magnetic field structure of the emission sites and put additional constraints on the emission models.
Recently, changes in the polarization (amplitude and angle) have been used to investigate the origin of  the $\gamma$-ray emission in several blazars \citep{marscher:2008a, marscher:2010a,abdo:2010:fermi:3c279:opticalpolarimetry,barresdealmeida:2010a}. The observed pattern -- a change in the HE/VHE flux coinciding with a rotation in the electric vector position angle (EVPA) and/or a change in the relative polarization -- could result from: a non-axisymmetric magnetic field through which the emission region propagates, a swing of the jet across the line of sight, or a curved trajectory of the emission region e.g. following helical magnetic field lines or a bent jet \citep{abdo:2010:fermi:3c279:opticalpolarimetry}.

\paragraph{\HST 2001-2008}
Regular optical polarimetry observations of the \M jet have been performed with the \HST from 2004 to 2006 (Sec.~\ref{Sec:DataOpticalLT}; \citealt{perlman:2011a}). The measured degree of polarization and the polarization angle for the core region and HST-1 are shown in Fig.~\ref{Fig:OpticalPolarimetry}, left panel. While both components show variable polarization their overall behavior is very different: during the flaring period (2004-2007) HST-1 shows a clear correlation of the fractional polarization with the total flux, with the fractional polarization ranging from 20 to 40\,\%. The EVPA remains almost constant $\sim7-8$\degree off the direction of the jet ($-69.5$\degree; gray line in Fig.~\ref{Fig:OpticalPolarimetry}, lower panel). The nucleus, on the other hand, displays a highly variable EVPA with a lower overall polarization of 2 to 10\,\%. The EVPA changes are not correlated with the total flux or the fractional polarization and appear, within the given sampling, erratic. The \HST polarimetry data covers the 2005 VHE flare but, apart from the previously mentioned correlation of the HST-1 photometric flux and fractional polarization, no correlations connected to the 2005 VHE flaring episode are apparent from the data.

\paragraph{LT 2010}
Triggered by the detection of the 2010 VHE flare the LT started to take regular optical polarimetry observations of \M in April 2010. The resolution of the instrument is not sufficient to separate the core and HST-1 and, therefore, one value for the combined core + HST-1 region is given.
The LT detects clear variability of the degree of polarization of the core + HST-1 complex and marginal evidence ($\sim2\sigma$) for variability of the EVPA (Fig.~\ref{Fig:OpticalPolarimetry}, right panel). The polarization degree changes between $\sim2$ and $\sim12$\,\% but without any apparent correlation with the VHE $\gamma$-ray flux.
The measured EVPA indicates changes at the time of the VHE flaring episode from $\sim -130$\degree to $\sim-60$\degree ($\sim 70$\degree in 3 days) and later settles to a stable EVPA of $\sim -20$\degree about 50\,days after the flare, though the only marginal indication for EVPA variability and the lack of continuous monitoring before and directly after the VHE flare makes it difficult to interpret these changes in terms of a continuous rotation of the EVPA. 
The observed polarization variability pattern is compatible with the pattern previously observed by \HST for the core region and different to what has been observed from HST-1. This indicates that the 2010 LT data is possibly dominated by emission from the core.


\section{VHE flares \& MWL correlations}\label{Sec:VHEFlaresVsMWL}

Over the last 10 years \M has been extensively monitored all across the electro-magnetic spectrum from radio to VHE (Fig.~\ref{Fig:MWL-longterm-lightcurve}). This large data-set can be used to investigate MWL correlations and thereby probe the origin of the VHE $\gamma$-ray emission. In principle, many different physics processes could contribute to the production of VHE $\gamma$-ray emission in \M, e.g. annihilation of massive dark matter particles \citep{baltz:2000a}, cosmic ray interactions in the extended radio lobes and the surrounding cluster \citep{pfrommer:2003a}, particle acceleration in the relativistic jets, etc. The detected short-term variability with timescales of the order of days and the limits on the location place strong constraints on possible scenarios \citep{aharonian:2006:hess:m87:science}, leaving the inner jet and the close vicinity of the SMBH as the most probable emission sites. In the following, the MWL behavior of the two most prominent features in the innermost structure of \M, namely the HST-1 knot and the core, are discussed in the light of the VHE flaring activity.

\paragraph{HST-1}
Between 2001 and 2008 the first bright feature in the jet resolved by the \HST in the optical, HST-1, underwent a spectacular flare detected in radio, optical, and X-rays \citep{perlman:2003a,harris:2003a}. The flare displayed a relatively smooth rise over several years with a flux increase by more than a factor 50 in X-rays and optical. The flux peaked in the beginning of 2005 \citep{harris:2006a,harris:2009a} around the same time when the enhanced activity level and the first short term variability had been detected at VHE \citep{aharonian:2006:hess:m87:science}. HST-1 has been discussed as a possible site for the VHE $\gamma$-ray emission \citep[e.g.][]{stawarz:2006a,cheung:2007a,harris:2009a}. While the size of HST-1 as a whole is too large to account for the short-term variability detected at VHE (following causality arguments), high resolution VLBA radio observations resolve HST-1 into several, partially unresolved sub-structures \citep{cheung:2007a}. These sub-structures also display apparent superluminal motion up to $4.3\,c \pm 0.7\,c$ (see also \citealt{giovannini:2011a}).
In combination with the detected synchrotron X-ray emission and strong polarization of the radio-to-optical continuum, this indicates that efficient in-situ acceleration of the radiating particles is taking place in compact sub-volumes of the HST-1 region, characterized by well-organized magnetic field and relativistic bulk velocities.
On the other hand, during the 2008 and 2010 VHE flares HST-1 was in a low flux state without pronounced activity at radio or X-ray wavelengths, thus disfavoring it as the origin of the VHE $\gamma$-ray emission during these episodes.

\paragraph{Core}
The direct vicinity of the SMBH and the jet base have been proposed as possible production sites of the VHE $\gamma$-ray emission \citep[e.g.][]{reimer:2004a,ghisellini:2005a,tavecchio:2008a,neronov:2007a,rieger:2008a,lenain:2008a,barkov:2010a,giannios:2010a,levinson:2011a}.
\M is only a weak IR source ($\nu L_{\nu} \sim 10^{41}$\,erg/s;  \citealt{perlman:2001b})
and, therefore, VHE $\gamma$-rays are most likely able to escape even from the close vicinity of the SMBH without suffering strong absorption due to $\gamma \gamma$-interactions \citep{neronov:2007a,brodatzki:2011a}, although the debated origin of the observed IR photons and the poorly known structure of the accretion disk in the \M core, which both play a crucial role in this context, should be kept in mind (see the discussion in \citealt{cheung:2007a} and \citealt{li:2009a}).
The \M jet base has been imaged with VLBI with sub-mas resolution \citep[see e.g.][and references therein]{ly:2007a}. It shows a resolved, edge brightened structure to within 0.5\,mas of the core (0.04\,pc) and indications for a weak counter-jet feature, suggesting that the SMBH lies within the central beam of the VLBI observations. At even shorter distances to the core the jet has a wider opening angle, which is interpreted as the jet collimation zone \citep{junor:1999a}.

In 2008, densely sampled 43\,GHz radio observation of the innermost jet regions revealed a flare of the radio core (flux increase of $\sim30$\,\%; \citealt{acciari:2009b:m87joinedcampaign:science}). At the same time, a flare at VHE and a subsequent enhanced X-ray flux from the core region followed by a sharp decrease were detected. The observed MWL variability pattern supported the interpretation that the VHE $\gamma$-ray emission likely originates from the close vicinity of the SMBH near the jet base \citep{acciari:2009b:m87joinedcampaign:science}. The observed MWL behavior (VHE and radio flux, radio map) is well described by a simple, phenomenological model where the VHE flares are associated with the injection of plasma at the jet base (\citealt{acciari:2009b:m87joinedcampaign:science} supporting online material). As the injected plasma blobs travel down the jet, they expand and become transparent for radio emission leading to the observed radio feature.

In contrast, radio observations taken in 2010 contemporaneous with the VHE flare show no enhanced radio flux from the core region (Fig.~\ref{Fig:MWL-longterm-lightcurve}): VLBA 43\,GHz ToO observations triggered by the detection of the VHE flare indicate a stable flux state of the core (1.2\,mas) and the inner jet (1.2 to 5.3\,mas) at the previously detected flux levels. In addition, EVN 5\,GHz, VLBA 2.3\,GHz, and MOJAVE 15\,GHz measurements also show no indication of an enhanced radio flux state from the core in 2010. A direct comparison of these data with the 43\,GHz data is difficult given the difference in resolution and the missing overlap during the 2008 flare.\footnote{Noteworthy, one of the highest flux states in the MOJAVE 15\,GHz data is measured in 2008 shortly after the VLBA 43\,GHz observations ended.} In the optical, only a combined measurement for the core and HST-1 with limited sensitivity by the LT is available, which does not indicate any strong activity in the two components in 2010.

On the other hand, \Chandra X-ray observations of the core show an enhanced flux $\sim3$\,days after the peak of the VHE $\gamma$-ray emission in 2010 (see Fig.~\ref{Fig:MWL-longterm-lightcurve}). The flux is enhanced by a factor $\sim2$ for a single measurement and than drops back to a lower state less than two days later. The observed variability timescale is significantly shorter (by a factor $\sim10$) then the shortest X-ray variability measured previously from the \M core \citep[$\lesssim20$\,days;][]{harris:2009a}. Further details on the \Chandra X-ray data from 2010 will be reported in a companion paper \citep{harris:2011a}.

It should be noted that the X-ray fluxes measured from the core during the time of the VHE flares in 2008 and 2010 are the two highest measurements since the start of the \Chandra observations in 2002. This coincidence can be interpreted as indication that the VHE $\gamma$-ray emission in 2008 and 2010 originates from the X-ray core region. During the 2005 VHE flaring episode no enhanced X-ray emission from the core was detected. At that time, HST-1 was more than a factor 30 brighter than the core region in X-rays leading to uncertainties in the flux estimation of the core (e.g. 'eat-thy-neighbor' effect; see Appendix~A of \citealt{harris:2009a}) which could suppress the detection of a core flare.

\section{Discussion}\label{Sec:Discussion}

\paragraph{General considerations}
The VHE and MWL data in this paper can be interpreted in two fundamentally different ways: (1) each of the VHE flares detected originates from a different emission region and/or process, or (2) they have a common origin. Support for the former interpretation  comes from the difference in the VHE light curves (variability pattern and the overall duration of each episode), as well as from the difference in the apparent MWL correlations in the different years. In this interpretation, all previously derived models remain possible and an additional explanation for the 2010 flare has to be found. Observational support for the latter interpretation, involving a common origin of the VHE flares, follows from the similarities in the detected peak fluxes, flux doubling timescales, and also in the spectral shapes of the different VHE flares \citep[for details on the VHE spectrum see][]{aharonian:2006:hess:m87:science,albert:2008:magic:m87,aliu:2011a:veritas:m87:2010}. In this interpretation the only remaining MWL signature for the VHE flares is an X-ray flare of the core region (see previous section) locating the VHE $\gamma$-ray emission site in the central resolution element of the \Chandra observations ($0.6\,\arcsec \sim$ 50\,pc).\footnote{It should also be noted that direct observational limits on the extent of the VHE emitting region constrain its projected size to be $\lesssim 14$\,kpc centered on the radio core \citep{aharonian:2006:hess:m87:science}.}

In addition, it is not known whether the quiescent and flaring VHE $\gamma$-ray emission are produced in the same region and by the same process.
While, in principle, they can be produced in two distinct locations by two different processes, the similarities of the VHE $\gamma$-ray spectrum between the two states might suggest a common origin, or at least a very similar physical process involved. A common origin of the quiescent and flaring VHE fluxes is therefore anticipated below, even though different emission regions dominating the two states cannot be excluded.
Interestingly, the HE spectrum measured by \FermiLAT joins smoothly with the VHE spectra derived for the quiescence state \citep{abdo:2009:fermi:lat:m87} possibly indicating a common origin and, therefore, HE-VHE flux correlations could be expected. Unfortunately, due to the low flux of \M in the HE band and the limited sensitivity of \FermiLAT, no conclusive statement on such correlation can be derived from the data presented in this paper.

In the following, the discussion will mainly focus on the case where a common origin of all VHE flares in the X-ray core ($0.6\,\arcsec \sim$ 50\,pc) associated with an X-ray flare is assumed.

\paragraph{Characteristics of the VHE flares}
The VHE flares are characterized by a variability timescale $t_{\rm var} \simeq 1$\,day, and a broad band power law spectrum extending from $\sim$0.1\,TeV up to at least 10\,TeV with photon index $\Gamma \simeq 2.3$, weakly variable during flares \citep{aharonian:2006:hess:m87:science,albert:2008:magic:m87,aliu:2011a:veritas:m87:2010}.
The observed VHE flux during the flaring episodes reached similar maximum flux levels of $\Phi_{>0.35\,{\rm TeV}} \simeq (1-3) \times 10^{-11}$\,ph\,cm$^{-2}$\,s$^{-1}$. With the adopted distance of $d=16.7$\,Mpc and using the average photon index $\Gamma=2.3$, the corresponding isotropic VHE luminosity is $L_{\rm VHE} = 4 \pi d^2 \, \Phi_{>E_0} \, E_0 \, (\Gamma -1)/(\Gamma-2) \simeq (0.8-2.4) \times 10^{42}$\,erg\,s$^{-1}$. If extrapolated down to the HE range (0.1\,GeV) with the same photon index, the total $\gamma$-ray luminosity of the flaring events would be even higher, namely $L_{\rm HE-VHE} \sim 10^{43}$\,erg\,s$^{-1}$. This is a non-negligible amount given that the bolometric accretion luminosity of the \M nucleus is relatively low, $L_{\rm acc} \sim 10^{42}$\,erg\,s$^{-1}$ \citep{reynolds:1996a,dimatteo:2003a,kharb:2004a}, and the total kinetic luminosity of the jet is also quite modest, $L_{\rm j} \sim 10^{44}$\,erg\,s$^{-1}$ \citep{bicknell:1996a,owen:2000a}. The efficiency of the VHE $\gamma$-ray production in \M is therefore an issue, even taking into account order-of-magnitude dimmer quiescence fluxes. More specifically, the observed VHE luminosity during the flaring events \citep[which is approximately equal to the average/quiescence bolometric $\gamma$-ray luminosity in both HE and VHE bands; see][]{abdo:2009:fermi:lat:m87} is of the order of the accretion luminosity in the \M system, constituting at the same time about $1\%$ of the jet total kinetic luminosity, $L_{\rm VHE} \sim L_{\rm acc} \sim 0.01 \times L_{\rm j}$.

Interestingly, assuming the observed emission is moderately beamed with a Doppler factor $\delta$ of the order of a few and that, when viewed at smaller inclinations, the beaming of the nuclear jet in the \M system would be the same as the one deduced for blazar sources, namely $\delta^{\star} \simeq 10-30$, the isotropic VHE flaring luminosity of \M observed at smaller viewing angles would read $L_{\rm VHE}^{\star} \simeq (\delta^{\star} / \delta)^4 \, L_{\rm VHE} \simeq (10^{44}-10^{46})$\,erg\,s$^{-1}$. This is in the range of VHE flaring luminosities of blazars of the BL Lac type, for which low-power radio galaxies like \M are believed to constitute a parent population \citep{urry:1995a}.
Moreover, in such a scenario the observed variability timescales are scaled down by the ratio of the Doppler factors $\Delta t^\star =  (\delta / \delta^{\star}) \, \Delta t \sim (0.5 - 0.2) \, \Delta t$ becoming of the order of a few hours, which is less than the timescale derived from the size of the Schwarzschild radii via causality arguments ($\sim$day). Variability timescales shorter then the causality timescales are also observed in flaring VHE blazars, where, in some cases, variability timescales down to a few minutes imply even higher Doppler factors of $O(100)$ \citep[e.g.][]{albert:2007:magic:mkn501, wagner:2008a, abramowski:2010:hess:pks2155:timing}.
This agreement can be considered as support for a blazar-like origin of the $\gamma$-ray emission in \M. The VHE flux changes by a factor of $10$ and
a VHE flare duty cycle of $O(10\%)$, as estimated in this paper, would also be consistent with such a hypothesis \citep{wagner:2008a}.
Note in this context that with relativistic beaming involved the power emitted during the VHE flares would be reduced with respect to the isotropic luminosity estimated above as $L_{\rm em,\,VHE} \simeq \Gamma_j^{-2} L_{\rm VHE}$, where $\Gamma_j$ is the bulk Lorentz factor of the emitting region.

Particularly interesting and constraining are the distinct timing properties of the one-day-long VHE flares of \M, recognized clearly and quantified for the first time in this paper. These include repetitive outbursts (2005), erratic flux changes (2008), exponential flux increases/decays for well-defined isolated events, and significantly different rise and decay times (2010). If indeed the three flares share a common origin, every model dealing with the VHE $\gamma$-ray emission of M87 has to be able to accommodate such a diversity in the observed variability patterns.

Lastly, MWL correlations (or the apparent lack of such) have to be addressed as well, meaning in particular the emerging connection between the VHE and X-ray bands with no accompanying flux variations at radio or optical frequencies.  It should be noted, however, that unlike the order-of-magnitude flux increases observed in the VHE regime, the observed variability in the X-ray regime is of a much lower amplitude (flux changes by a factor of $\sim 2$ only), corresponding to a rather moderate X-ray core peak luminosity of the order of $\sim 10^{41}$\,erg\,s$^{-1}$. This value has to be taken with some caution though since, up to now, there are no truly simultaneous X-ray core observations during a VHE flare (i.e. during the night of the peak of the VHE $\gamma$-ray emission).

In the following paragraphs, existing models for VHE $\gamma$-ray emission from the \M core are briefly reviewed and discussed in the light of the new observational result presented in this paper. The majority of models discussed have been published before the detection of \M by \FermiLAT \citep{abdo:2009:fermi:lat:m87}. Models published before 2006 are also not constrained by the VHE short-term variability discovered by \He \citep{aharonian:2006:hess:m87:science}.

\paragraph{Observations versus modeling}

Let us start with the 'misaligned blazar'-type, or rather 'misaligned BL Lac'-type scenarios \citep[noting at the same time that the standard leptonic 'homogeneous one-zone SSC' approach has been often considered to fail in explaining the observed VHE properties of the \M nucleus; but see also the discussion in][]{abdo:2009:fermi:lat:m87}. One of the first models of this kind was discussed by \citet{reimer:2004a}, and involved a mixture of hadronic and leptonic emission processes operating within its relativistic outflow at distances from the jet base small enough to assure $R<   c \, t_{\rm var} \delta$ for the emission region size $R \sim 10^{16}$\,cm, variability timescale $t_{\rm var} \sim 1$\,day, and the expected moderate beaming, $\delta \sim$a~few. Designed to account for the non-simultaneous data known at that time, the HE/VHE continuum was dominated by the synchrotron radiation of protons and muons in strong magnetic fields. While proton synchrotron radiation, with a maximum intrinsic cut-off at $\sim 0.3$\,TeV, appears unable to account for the measured VHE spectra detected up to $\sim10$\,TeV, synchrotron radiation from muons (produced in charged pion decays) and the pion cascade components could potentially extend the VHE spectrum to higher energies. This would then imply pion-production losses to be at least of the same order of magnitude as proton synchrotron losses. The low energy, millimeter--to--UV part of the spectrum is ascribed to the synchrotron emission of primary electrons with the same injection spectral index as the primary protons, and the high, non-simultaneously measured X-ray flux considered to be dominated by an additional component. A direct VHE/X-ray correlation with a high X-ray flux level is thus not expected in such a scenario.

\citet{georganopoulos:2005a} discussed a leptonic version of the blazar-type modeling of the \M nucleus, relaxing the assumption regarding the homogeneity of the emission region. More precisely, \citeauthor{georganopoulos:2005a} considered the case of a relativistic jet decelerating substantially on sub-pc distances from the core, and showed that the velocity difference between a faster and a slower portion of the outflow could lead to the enhanced inverse-Compton emission of the former one in the TeV range (due to a relativistic boost of the energy density of soft photons produced in one portion of the flow in the rest frame of the other), at a level allowing a fit to the VHE data. In the model, the core emission observed at longer wavelengths (including the GeV range) was predominantly due to the slower and outer parts of the decelerating jet. A related scenario was analyzed by \citet{tavecchio:2008a}, who assumed instead radial velocity stratification, involving a relativistic jet spine surrounded by a slower sheath. In the particular model fits presented by \citeauthor{tavecchio:2008a}, the observed radio--to--HE emission of the unresolved \M core was dominated by the radiative output of the jet spine, while the observed VHE $\gamma$-ray emission was produced mainly in the jet boundary layer. Both models are characterized by compact sizes of the VHE $\gamma$-ray emission region ($R < 10^{17}$\,cm), moderate beaming, and no obvious --- if any --- correlation between the VHE and lower-frequency bands. Importantly, the VHE spectra calculated in the framework of both models were rather soft, due to the Klein-Nishina and $\gamma$-ray opacity effects involved \citep[see the discussion in][]{tavecchio:2008a}. Multi-zone approaches, meaning a doubled number of the model free parameters, could possibly allow for some modifications to the presented fits enabling one to accommodate some additional MWL constraints (for example the VHE/X-ray correlation), but the observed flat flaring spectra of \M in the VHE range will remain problematic in both cases.

Blazar type models assuming a different structure of the emission region are discussed in \citet{lenain:2008a} and \citet{giannios:2010a}. In those, the bulk of the observed emission of the unresolved \M jet was proposed to be produced in extremely compact sub-volumes of the main outflow. In the scenario analyzed by \citeauthor{lenain:2008a}, such compact sub-volumes were assumed to be multiple blobs ejected from the core with modest or highly relativistic bulk velocities into a cone with a large opening angle. In the framework of the model considered by \citeauthor{giannios:2010a}, on the other hand, one is dealing with fast mini-jets moving in random directions within a relativistic `large-scale' outflow. Complex setups of the models resulted in the fact that different variability patterns and various variability timescales could, in principle, be expected, depending on a particular choice of the model free parameters (for which broad ranges of values could be considered). In the particular fits presented by \citet{lenain:2008a}, the authors considered for example the radio--to--UV continuum of the \M core to be produced by the extended though still unresolved portion of the jet, and argued that the synchrotron and inverse-Compton emission of tiny blobs moving within such a jet may account for the observed nuclear X-ray and $\gamma$-ray fluxes, respectively, including not only the quiescent but also flaring states. The anticipated VHE/X-ray correlation with no accompanying flux variations at longer wavelengths is thus an interesting feature of the model.

Yet the crucial assumption regarding linear sizes of the emitting blobs significantly smaller than the Schwarzschild radius of the \M black hole, $R \ll R_{\rm S}$, might be regarded as questionable \citep[see the discussion in][]{begelmann:2008a}. 
A possible physical justification for the formation of such ultra-compact sites of the enhanced energy dissipation close to SMBHs in AGN jets, and \M in particular, was given by \citet{giannios:2010a}, who speculated that this is the magnetic reconnection process which may trigger compact beams of plasma (`mini-jets'), which then propagate with relativistic bulk velocities within a strongly magnetized, extended and similarly relativistic outflow. In their application to the specific case of \M, \citet{giannios:2010a} demonstrated that the synchrotron and inverse-Compton emission of the reconnection-driven mini-jets could account for the observed X-ray and VHE nuclear fluxes, somewhat in analogy to the results obtained by \citet{lenain:2008a}, again under particular model assumptions regarding the model free parameters. In the end, not only the fast VHE variability and the flat VHE spectra, but also the VHE/X-ray correlation could be successfully accommodated in the model.

Possibly challenging for scenarios involving ultra-compact emission regions is also the required efficiency of the VHE $\gamma$-ray production. As discussed above, the VHE flares observed in the \M system are characterized by an isotropic luminosity constituting at least $1\%$ of the total kinetic power of the \M jet. Whether such a power can indeed be efficiently dissipated within tiny sub-volumes of the jet remains an open question, even taking into account the expected beaming corrections.

Magnetospheric particle acceleration and emission models have also been invoked to explain the observed VHE emission in \M \citep[see e.g.][for a review]{rieger:2011a}. Assumed to be operating close to the event horizon of the central supermassive black hole, the anticipated variability timescale (light-crossing argument) could be as small $t \simeq 2 G M_{\rm BH} / c^3 \simeq 0.35$\,day, and thereby satisfy the observed variability constraints. The observed radio-VHE correlation during the 2008 VHE flare has been interpreted to provide additional support for such an approach \citep{acciari:2009b:m87joinedcampaign:science}. Usually, efficient particle acceleration in these scenarios is related either to gap-type \citep{neronov:2007a,levinson:2011a} or centrifugal-type processes \citep{rieger:2008a} occurring in the magnetosphere around a rotating black hole. While the former can be very efficient and lead to the onset of an electromagnetic pair cascade (triggered by the absorption of ambient photons up-scattered to high energies by electrons accelerating in the gap), the latter is less efficient such that the VHE spectrum is expected to be shaped by the ambient soft photon spectrum \citep[see also][]{vincent:2010a}. Both approaches appear to be able to reproduce the observed hard VHE flaring characteristics, but may need an additional contribution to account for the HE continuum.
     
Recently, a different type of modeling has been brought forward to explain the observed VHE $\gamma$-ray emission from \M: \citet{barkov:2010a} argued that interactions of a relativistic and strongly magnetized outflow around the jet formation zone with a star partially tidally disrupted by the interaction with the SMBH can lead to a very efficient production of hadronic-originating VHE photons \cite[see also][for a different version of the jet-star interaction model for blazar-type sources]{bednarek:1997a}. The predicted exponential character of the flux increase and decay during the resulting VHE event, a flat GeV to TeV $\gamma$-ray spectrum from proton-proton interactions, as well as the accompanying X-ray flux enhancement due to the free-free emission of the shocked cloud of the stellar matter, are particularly interesting features of the model which should be kept in mind. Other  scenarios could yet be considered in the same context as well, involving stochastic acceleration of high energy particles within a turbulent accretion flow close to the event horizon of a SMBH \citep[in analogy to the model proposed by][in the context of VHE $\gamma$-ray emission of Sgr A*]{liu:2006a}, or reconnection-driven impulsive acceleration of electron-positron pairs in a magnetized corona of the accretion disk, e.g., in analogy to the model proposed by \citet{zdziarski:2009a} in the context of VHE observations of the Galactic X-ray binary system Cygnus X-1 \citep[see also][]{degouveiadalpino:2010a}.

\paragraph{HST-1}
The arguments against association of the detected VHE $\gamma$-ray emission with the non-thermal activity of the HST-1 knot are twofold: First, rapid variability of the VHE continuum involving day-long flares implies that the emission region size is significantly smaller than the inferred size of the knot. Second, apart from the 2005 event, no correlation between the VHE activity and the radio--to--X-ray synchrotron radiation of the HST-1 flaring region, variable on the timescales of weeks and months, has been established. The first argument has to be taken with caution, though. That is because the HST-1 flaring region is basically unresolved even for radio interferometers, as already noted before. More importantly, very recently it has been found that rapidly variable high-energy radiation can be generated far away from the central jet engine, within the outer parts of relativistic outflows. In particular, the analysis of the $\gamma$-ray emission of a bright quasar PKS\,1222+216 indicates that the day-long VHE flares in this object originate most likely in compact sub-volumes of a relativistic jet at parsec distances from the active center \citep{aleksic:2011:magic:pks1222,tanaka:2011a,tavecchio:2011a}.
Interestingly, analogous phenomena seem to occur on even larger scales as well. For example, \Chandra monitoring of the Pictor\,A radio galaxy reveals that the synchrotron X-ray emission of the knots located at tens of kpc from the jet base is variable on the timescales of years. This is much shorter than the variability timescale expected following the causality arguments for a given jet radius at the position of the knot, which is of the order of thousands of years \citep{marshall:2010a}.
Hence, it remains formally possible that also in the case of the HST-1 knot, located about 100\,pc from the \M nucleus, rapid VHE flares are being generated on timescales shorter than the ones expected for a homogeneous outflow.

Yet, the indications found in this paper for correlation between the VHE flares and an X-ray flux increases of the nucleus of \M provide observational support for the idea that the observed VHE $\gamma$-ray emission, at least during the flaring episodes, is associated with the innermost parts of the jet or with the closest vicinities of the SMBH in the center of this radio galaxy.
Again, in principal, it remains plausible that, while the unresolved core dominates the flaring states, the quiescent VHE $\gamma$-ray emission of \M is instead related to the HST-1 knot. However, above we have advocated for a common origin of the quiescence and flaring VHE fluxes based on the spectral similarity of the two states. If the favored hypothesis is correct indeed, then interesting constraints on the physical parameters of HST-1 can be derived. That is because the inverse-Compton up-scattering of ambient photon fields to the VHE range by the high-energy electrons producing the observed radio--to--X-ray synchrotron radiation of the knot is at some level inevitable, with the efficiency depending most crucially on the unknown magnetic field intensity within the considered jet region. Since for the equipartition value of the jet magnetic field a relatively intense VHE $\gamma$-ray emission of HST-1 should be expected \citep{stawarz:2006a,cheung:2007a}, the upper limits for the radiative output of the knot in $\gamma$-rays (following from the fact that the observed VHE flux is associated with the nucleus) implies a strong, possibly even dynamically relevant jet magnetic field at hundreds-of-pc distances from the jet base. This should be then considered as an important support for the MHD models of AGN jets, like the one presented in the particular context of the \M core and HST-1 knot by \citep{nakamura:2010a}, and discussed further from the observational perspective by \citet{perlman:2011a} and \citet{chen:2011a}. Analogous constraints emerging from the VHE data can be investigated for the outer parts of the \M jet as well \citep{stawarz:2005a,hardcastle:2011a}.

\section{Summary \& Conclusions}\label{Sec:SummaryConclusions}

During a joint VHE monitoring campaign on the nearby radio galaxy \M in 2010, a major flux outburst was detected, triggering further VHE and MWL observations. The coordinated observations led to the best-sampled VHE light curve during a flaring state from this source (21 observations in 15\,days), revealing a single, isolated outburst. The measured VHE light curve of the flare is well described by a two-sided exponential function with significantly different flux doubling times of $\tau_\text{d}^{\text{rise}} = (1.69 \pm 0.30)$\,days for the rising and $\tau_\text{d}^{\text{decay}} = (0.611 \pm 0.080)$\,days for the decaying part, i.e. a difference of a factor $2.77 \pm 0.62$. This measurement provides the shortest and the most precisely determined VHE variability timescale of \M today, and the first detection of a significantly asymmetric VHE flare profile in the source.

In comparison to previous VHE flares detected in 2005 and 2008, the 2010 flare shows similar timescales and peak flux levels, but the overall variability pattern is somewhat different from the more extended periods of flaring activity with several flux maxima observed before, though the statistics and the sampling of the previous VHE flares limit a definitive conclusion on that matter. From the VHE long-term light curve the duty cycle for VHE flares is estimated to be $< 4-28$\%, depending on the assumed threshold flux defining a VHE high state.

VLBA 43\,GHz observations, triggered by the detection of the VHE flare, show no indications for an enhanced radio emission from the jet base in 2010. This is in contrast to observations in 2008, where the detection of a radio outburst of the core contemporaneous with the VHE flare lead to the conclusion that the VHE $\gamma$-ray emission is likely produced in the direct vicinity of the SMBH \citep{acciari:2009b:m87joinedcampaign:science}.
\Chandra X-ray observations, taken $\sim3$\,days after the peak of the VHE $\gamma$-ray emission, show a high flux state of the core region in 2010, supporting the interpretation that the VHE flare originates from the innermost jet regions.

The long term (2001-2010) light curve of \M, spanning from radio to VHE, is investigated for a common MWL signature accompanying the three VHE flares.
No unique signature is found.
Observations of the jet component HST-1, which has also been proposed as a possible emission site of the VHE $\gamma$-ray emission, show no enhanced activity in 2008 and 2010, disfavoring HST-1 as the origin of the VHE flares during those years.
In 2008 and 2010 the VHE flares are accompanied by a high state of the X-ray core with a flux increase by a factor $\sim2$, while in 2005 the strong flux dominance (more than a factor 30) of the nearby X-ray feature HST-1 could have suppressed the detection of such an increase of the core emission. Associating the VHE flares with the X-ray flares from the core places the emission site in the central resolution element of \Chandra ($0.6\,\arcsec \sim$ 50\,pc).

Several models have been proposed to explain the observed VHE $\gamma$-ray emission from \M, most of which were based on the `misaligned BL Lac' hypothesis. And, in fact, several observed properties of the source, including the broad-band $\gamma$-ray spectrum, the multi-wavelength character of the VHE flares, and their overall energetics, could be considered as support for the blazar-like models for the VHE $\gamma$-ray emission.
Yet, the particular one-zone or two-zone emission models proposed involving only moderate beaming, both hadronic and leptonic, have difficulties in explaining VHE/X-ray correlated variability, for which indications are found in this paper, as well as the relatively flat VHE spectrum of \M (photon index $\Gamma \sim 2.3$) extending up to $\sim 10$\,TeV.
More complex blazar-type scenarios, involving ultra-compact emission regions and more substantial beaming, may, instead, face difficulties in accounting for the flares' energetics with the isotropic peak luminosity during the flares reaching $L_{\rm VHE} \simeq (0.8-2.4) \times 10^{42}$\,erg\,s$^{-1}$, which is of order of the accretion luminosity and about 1\% of the total jet power. Other non-blazar emission models discussed in the context of the VHE observations of \M, which are quite successful in reproducing some of the established properties of the source, seem, on the other hand, challenged by the broad-band character of the $\gamma$-ray continuum, as long as the GeV photon energy range constrained by the \FermiLAT observations is included. All in all, even though no emission model could be identified as the most plausible one, important limitations for most of them have been identified and novel observational constraints have been presented in this paper. Thus, the case of \M demonstrates the relevance of systematic, long-term and multi-wavelength monitoring of nearby radio galaxies in understanding the origin of the high-energy emission of radio-loud AGN.


{
\acknowledgments
\small
The authors thank the anonymous referee for helpful suggestions which improved the manuscript.\\
The \He Collaboration acknowledges support of the Namibian authorities and of the University of Namibia
in facilitating the construction and operation of H.E.S.S., as is the support by the German Ministry for Education and
Research (BMBF), the Max Planck Society, the French Ministry for Research,
the CNRS-IN2P3 and the Astroparticle Interdisciplinary Programme of the
CNRS, the U.K. Science and Technology Facilities Council (STFC),
the IPNP of the Charles University, the Polish Ministry of Science and 
Higher Education, the South African Department of
Science and Technology and National Research Foundation, and by the
University of Namibia. We appreciate the excellent work of the technical
support staff in Berlin, Durham, Hamburg, Heidelberg, Palaiseau, Paris,
Saclay, and in Namibia in the construction and operation of the
equipment.\\
The \Ma Collaboration would like to thank the Instituto de Astrof\'{\i}sica de
Canarias for the excellent working conditions at the
Observatorio del Roque de los Muchachos in La Palma.
The support of the German BMBF and MPG, the Italian INFN, 
the Swiss National Fund SNF, and the Spanish MICINN is 
gratefully acknowledged. This work was also supported by 
the Marie Curie program, by the CPAN CSD2007-00042 and MultiDark
CSD2009-00064 projects of the Spanish Consolider-Ingenio 2010
programme, by grant DO02-353 of the Bulgarian NSF, by grant 127740 of 
the Academy of Finland, by the YIP of the Helmholtz Gemeinschaft, 
by the DFG Cluster of Excellence ``Origin and Structure of the 
Universe'', by the DFG Collaborative Research Centers SFB823/C4 and SFB876/C3,
and by the Polish MNiSzW grant 745/N-HESS-MAGIC/2010/0.\\
The VERITAS Collaboration acknowledges support from the US Department of Energy Office of Science, the US National Science Foundation, and the Smithsonian Institution, from NSERC in Canada, from Science Foundation Ireland (SFI 10/RFP/AST2748), and from STFC in the UK. We acknowledge the excellent work of the technical support staff at the FLWO and at the collaborating institutions in the construction and operation of the instrument.\\
The \textit{Fermi} LAT Collaboration acknowledges generous ongoing support from a number of agencies and institutes that have supported both the development and the operation of the LAT as well as scientific data analysis. These include the National Aeronautics and Space Administration and the Department of Energy in the United States, the Commissariat \`a l'Energie Atomique and the Centre National de la Recherche Scientifique / Institut National de Physique Nucl\'eaire et de Physique des Particules in France, the Agenzia Spaziale Italiana and the Istituto Nazionale di Fisica Nucleare in Italy, the Ministry of Education, Culture, Sports, Science and Technology (MEXT), High Energy Accelerator Research Organization (KEK) and Japan Aerospace Exploration Agency (JAXA) in Japan, and the K.~A.~Wallenberg Foundation, the Swedish Research Council and the Swedish National Space Board in Sweden. Additional support for science analysis during the operations phase is gratefully acknowledged from the Istituto Nazionale di Astrofisica in Italy and the Centre National d'\'Etudes Spatiales in France.\\
Analysis of the Chandra data was supported by NASA grant GO0-11120X. \\
The Very Long Baseline Array is operated by the National Radio Astronomy Observatory, a facility of the NSF, operated under cooperative agreement by Associated Universities, Inc.\\
The European VLBI Network is a joint facility of European, Chinese, South African and other radio astronomy institutes funded by their national research councils. This effort is supported by the European Community Framework Programme 7, Advanced Radio Astronomy in Europe, grant agreement No. 227290.\\
This research has made use of data from the MOJAVE database that is maintained by the MOJAVE team \citep{lister:2009a}.\\
This research has made use of NASA's Astrophysics Data System
and of the python modules numpy, scipy, \& matplotlib \citep{hunter:2007a:matplotlib}.
}

\appendix
\section{Quantifying the difference between the 2008 and the 2010 VHE flares}
\label{Appendix:2008vs2010VHEFlare}

\begin{figure*}
\centering
\includegraphics[width=0.9\textwidth]{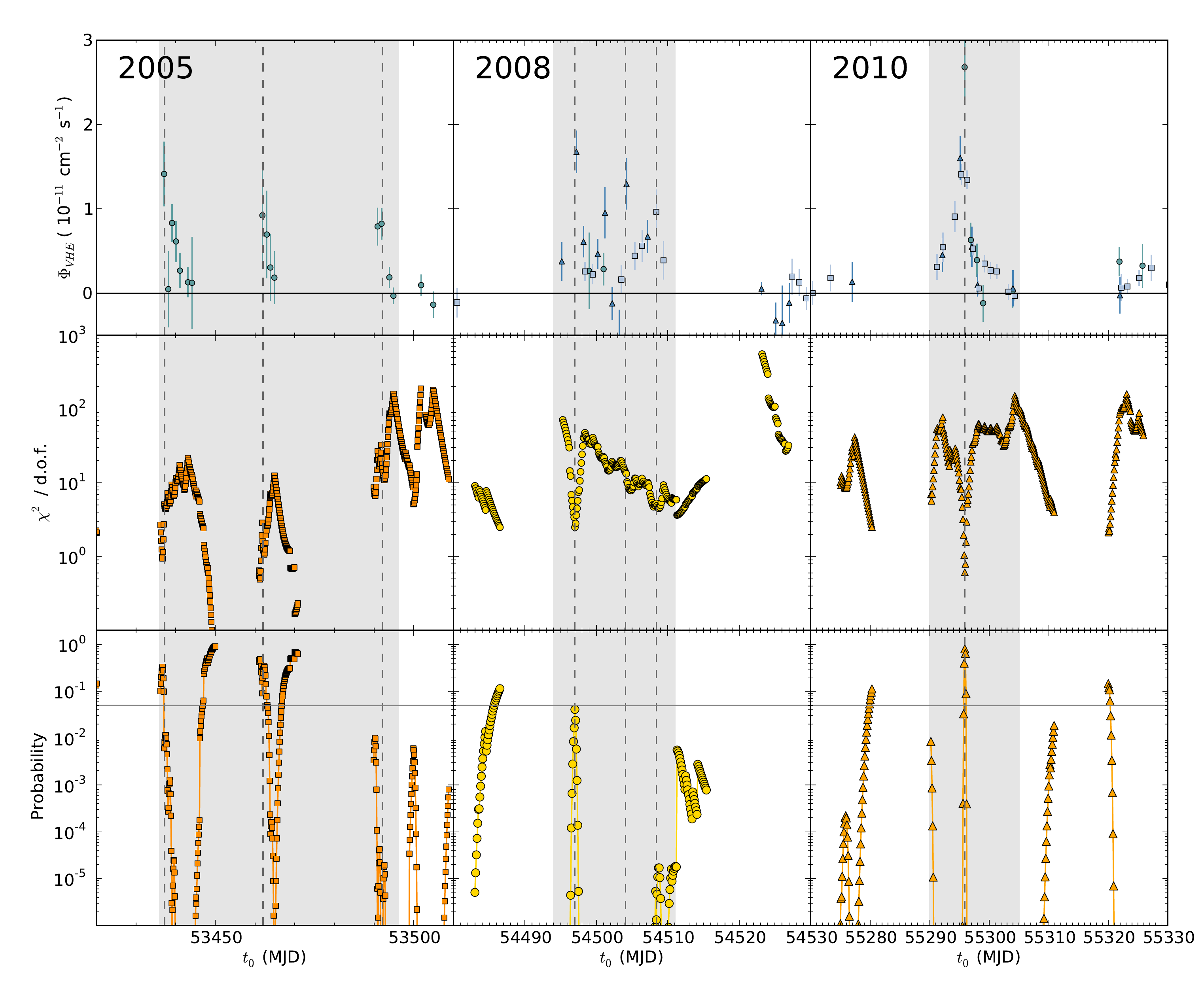}
\caption{VHE light curve (top), reduced chi-square $\chi^2 / \text{d.o.f.}$ (degree of freedom) (middle), and probability (bottom) vs the position of the peak $t_0$ of the best fit function to the 2010 VHE data for the 2005 (left), 2008 (middle), and 2010 VHE data (right). Values are calculated for data points inside a time-span of $t_0 - 7 < t_\text{data} < t_0 + 2$. The gray bands mark the same time period as in Fig.~\ref{Fig:MWL-longterm-lightcurve} and \ref{Fig:VHELightcurves2005-10}, denoting periods of increased VHE activity. Dashed vertical lines mark the positions of VHE flares. The horizontal line in the bottom panels marks a probability of 5\%.}
\label{Fig:2008vs2010VHEflare}
\end{figure*}

To investigate the compatibility of the 2008 and 2005 VHE data with the flare detected in 2010, the best fit exponential function to the 2010 data (Fig.~\ref{Fig:VHELightcurve2010}; Tab.~\ref{Table:FitFlareExp}) is taken as a template and compared with the 2008 and 2005 light curves. The position of the peak $t_0$ is varied (all other parameters fixed) and the resulting $\chi^2 / \text{d.o.f.}$ and probabilities are calculated for the data within a time span of length of the original fit length: $t_0 - 7 < t_\text{data} < t_0 + 2$. The results for the 2010, 2008, and 2005 data around the flaring episodes are shown in Fig.~\ref{Fig:2008vs2010VHEflare}, with the gray band again denoting the observation period with the flaring episode in each year. The statistics in 2005 are low resulting in several good probabilities (high probabilities $> 0.1$) over the flaring episode. For most parts of the 2008 flare, the probability for the 2010 fit describing the data is well below 1\%. The only notable exception is the beginning of the episode around MJD\,54496 where the first two data points are reasonably well described by the function resulting in a probability of $O(5\%)$. Note that the peak flux from the fit ($\sim 2 \times 10^{-11}\,\text{cm}^{-2}\text{s}^{-1}$) is similar to the peak fluxes measured in 2008. Varying the fit parameters in their errors does not significantly change the results. For comparison, the same study is shown for the 2010 data where, at the best fit position from the fit, the probability is $\sim$70\%.



\bibliography{m87_2010_mwl_vhe_v0.7.5.bbl}

\end{document}